\documentclass[aps,superscriptaddress]{revtex4}
\usepackage{amssymb,amsmath,epsfig}
\usepackage{subfigure}
\usepackage{color}
\usepackage[colorlinks=true, pdfstartview=FitV, linkcolor=blue, citecolor=red, urlcolor=magenta, breaklinks=true]{hyperref}
\usepackage{graphicx}
\usepackage{latexsym}
\usepackage{amsmath}
\usepackage{amsfonts}
\usepackage{amssymb}
\usepackage{booktabs}
\usepackage{multirow}
\usepackage{verbatim}
\graphicspath{{Fig/}}
\usepackage{orcidlink}

\begin{document}

\title{Absorption and quasinormal modes by rotating acoustic black holes in Lorentz-violating background}

\author{J. A. V. Campos \orcidlink{0000-0002-4252-2451}}\email{joseandrecampos@gmail.com}
\affiliation{Departamento de F\'{\i}sica, Universidade Federal de Campina Grande
Caixa Postal 10071, 58429-900 Campina Grande, Para\'{\i}ba, Brazil}

\author{M. A. Anacleto \orcidlink{0000-0003-4625-7322}}
\email{anacleto@df.ufcg.edu.br}
\affiliation{Departamento de F\'{\i}sica, Universidade Federal de Campina Grande
Caixa Postal 10071, 58429-900 Campina Grande, Para\'{\i}ba, Brazil}
\affiliation{Unidade Acad\^emica de Matem\'atica, Universidade Federal de Campina Grande
\\
58429-900 Campina Grande, Para\'{\i}ba, Brazil}

\author{F. A. Brito \orcidlink{0000-0001-9465-6868}}\email{fabrito@df.ufcg.edu.br}
\affiliation{Departamento de F\'{\i}sica, Universidade Federal de Campina Grande
Caixa Postal 10071, 58429-900 Campina Grande, Para\'{\i}ba, Brazil}
\affiliation{Unidade Acad\^emica de Matem\'atica, Universidade Federal de Campina Grande
\\
58429-900 Campina Grande, Para\'{\i}ba, Brazil}

\author{E. Passos \orcidlink{0000-0003-1718-6385}}\email{passos@df.ufcg.edu.br}
\affiliation{Departamento de F\'{\i}sica, Universidade Federal de Campina Grande
Caixa Postal 10071, 58429-900 Campina Grande, Para\'{\i}ba, Brazil}
\affiliation{Unidade Acad\^emica de Matem\'atica, Universidade Federal de Campina Grande
\\
58429-900 Campina Grande, Para\'{\i}ba, Brazil}

\author{Amilcar R. Queiroz \orcidlink{ 0000-0002-4785-5589}}\email{amilcarq@df.ufcg.edu.br}
\affiliation{Departamento de F\'{\i}sica, Universidade Federal de Campina Grande
Caixa Postal 10071, 58429-900 Campina Grande, Para\'{\i}ba, Brazil}

\begin{abstract}
In this work, we investigate the effects of Lorentz symmetry violation on the absorption cross section and quasinormal modes of a rotating acoustic black hole in (2+1) dimensions, within the regime of slow rotation and small Lorentz violating parameter $\alpha$. The absorption cross section was analyzed analytically, using the low and high frequency regimes, and numerically, through integration of the radial equation. The results showed that, in this regime, Lorentz violation increases the absorption cross section at all energy scales, with a contribution from the rotation parameter $B$ appearing even in the low frequency regime. For the quasinormal modes, we observed that symmetry breaking decreases the real part of the frequencies and increases the magnitude of the corresponding imaginary part, indicating a faster damping of the oscillations.
 
\end{abstract}
 \maketitle


\section{Introduction}
Black hole astrophysics has experienced remarkable progress in the last decade, driven by landmark experimental results. Highlighting is the first detection of gravitational waves by the LIGO-Virgo collaboration \cite{abbott2016tests, abbott2017gw170817}. Currently, with the addition of the \textit{Kamioka Gravitational Wave} Detector (KAGRA) laboratory in Japan, unprecedented improvements in precision have been achieved, enabling the detection of signals like GW250114, allowing the testing of black hole properties after collisions \cite{abac2025gw250114}, as well as studies on black hole spectroscopy \cite{abac2026black}. Other highly relevant experimental results include the images of the shadows of supermassive black holes obtained by the \textit{Event Horizon Telescope} (EHT) in the core of the galaxy M87 and Sagittarius A*, at the center of our own galaxy \cite{akiyama2019first, eventhorizon2019first, akiyama2022first}. New observations from EHT have revealed the inversion of magnetic fields in M87 \cite{akiyama2025horizon}.
One approach to investigating gravitational phenomena in controlled environments is the simulation of analog systems in the laboratory. The first acoustic black hole model was developed by Unruh in 1981 \cite{unruh1981experimental} to study fundamental properties such as Hawking radiation~\cite{Sakalli:2016mnk,Zhang:2011zzh,Anacleto:2022lnt,Anacleto:2023ali,Mondal:2025zuk}. In recent years, several analog gravity models have been developed \cite{Visser:1997ux,Ge:2010wx,anacleto2010acoustic,Anacleto:2011bv,Ge:2019our, Anacleto:2021nhm,barcelo2011analogue,Anacleto:2013esa}. In these models, the acoustic metric emerges when considering a moving fluid that reaches a local velocity greater than the speed of sound, creating a sonic horizon and consequently, an acoustic analog of a black hole.

Over the past few decades, a variety of theoretical studies and laboratory experiments have explored various aspects of the physics of analog black holes, providing experimental evidence of analog Hawking radiation \cite{weinfurtner2011measurement, steinhauer2016observation, munoz2019observation}, research on superradiance \cite{oliveira2010absorption, torres2017rotational, casadio2026quantum}, and investigations of quasinormal modes in analog systems \cite{cardoso2004quasinormal,Lepe:2004kv,Saavedra:2005ug, destounis2025vortices,Dolan:2010zza,Torres:2020tzs,Liu:2024vde}. Studies involving rotating systems have gained increasing attention. 
Recently, the spectra of scalar excitations propagating in rotating acoustic geometries were analyzed \cite{destounis2025vortices}.
Similarly, the introduction of acoustic metrics with frame-dragging effects (Lense-Thirring type) has enabled the investigation of phenomena such as acoustic shadows and the influence of rotation on the critical parameters of the system \cite{balali2025lense}.
In addition, acoustic shadow phenomena were also studied in~\cite{Ling:2021vgk,Guo:2020blq}

These experimental and theoretical advances establish a robust bridge between theoretical predictions and laboratory verifications.
In this work, we study the metric for an acoustic black hole in a scenario with Lorentz symmetry violation, obtained by incorporating terms that violate this symmetry into the Lagrangian of the Abelian Higgs model \cite{anacleto2010acoustic}. The presence of these terms modifies the equations that govern the fluid fluctuations, leading to several physical consequences. For example, the Hawking temperature associated with the acoustic horizon is directly affected by the Lorentz violation parameter \cite{anacleto2010acoustic}. In the case of rotating analogs, the violation term influences the superradiance phenomenon \cite{anacleto2011superresonance}. More recently, it has been found that the presence of Lorentz violation modifies the absorption and scattering cross section, as well as the quasinormal modes and acoustic shadows \cite{campos2024absorption}.

The motivation for deriving an acoustic metric from the Abelian Higgs model, which is originally formulated in the context of high-energy physics with the inclusion of a Lorentz symmetry-breaking term, lies in the possibility that, in extremely 
high-energy regimes, Lorentz violation effects may manifest themselves along with other phenomena, such as the formation of quark-gluon plasma (QGP). In this sense, it becomes particularly relevant to investigate the existence and properties of acoustic black holes in a QGP fluid subject to a Lorentz symmetry breaking. Studies on acoustic phenomena in QGP matter can be found in \cite{casalderrey2005conical, das2021hawking}, while the analysis of acoustic black holes in plasma fluids is discussed in \cite{de2008kerr, ditta2023particle}.
Another important application of analog models is the simulation of effects such as quasinormal modes originating from black holes. These quasinormal modes are fundamental to a better understanding of the properties of black holes. The study of perturbations in black holes began with the seminal work of Regge and Wheeler \cite{regge1957stability}, who investigated the stability of the Schwarzschild black hole. They identified that these perturbations evolve with a characteristic pattern of damped oscillations, where the frequency and damping time of these signals depend only on the parameters of the black hole, such as mass, charge, and angular momentum. Since then, several works have aimed to study the effects of quasinormal modes, including quantum corrections~\cite{Campos:2021sff, Anacleto:2021qoe, Yang:2022btw, Gingrich:2023fxu}, modified gravity models~\cite{Moulin:2019ekf, Chung:2024vaf} and black hole spectroscopy~\cite{Destounis:2021lum, Destounis:2023ruj, Lagos:2024ekd, Konoplya:2024lch}. Quasinormal modes in acoustic black holes can be found in both (2+1) and (3+1) dimensions. In particular, Cardoso et al. \cite{cardoso2004quasinormal} investigates the question of the dimensional instability of an acoustic black hole in the draining bathtub (DBT) model.

The study of spectral instabilities in analog systems has gained new momentum, with works investigating how the presence of vorticities in the fluid can perturb the effective potential and alter the spectrum of quasinormal modes \cite{correa2025black}, bringing the phenomenology of analogs even closer to astrophysical scenarios. Furthermore, it is also worth highlighting that there are significant advances in analytical methods, which allow a more precise and systematic description of the complex frequencies \cite{miyachi2025path}.
In the present work, we will examine the effects of Lorentz symmetry breaking in a rotating acoustic black hole metric in (2+1) dimensions, analyzing the behavior of the absorption cross section analytically in low and high frequency regimes and numerically for the entire spectrum. Our analysis is performed in the slow rotation regime and small values of the Lorentz breaking parameter $\alpha$. Applying the WKB approximation with higher-order corrections \cite{schutz1985black, iyer1987black, konoplya2003quasinormal}, we observe symmetry breaking in the quasinormal modes.

The paper is organized as follows. In Sec. \ref{S2}, we present the metric for an acoustic black hole with rotation in (2+1) dimensions with a term derived from the Lorentz symmetry breaking. In Sec. \ref{S3}, we analyze the differential absorption cross sections. We study the effects of the Lorentz violation term at low and high frequencies using the geodesic method and partial wave analysis for the acoustic metric. We extend the scattering study by verifying the results numerically. In Sec. \ref{S4} we introduce the study of quasinormal modes, verifying the behavior of the real and imaginary parts of the quasinormal frequency. Finally, in Sec. \ref{S5} we make our conclusions.

\section{The Lorentz Violating Model in an rotating acoustic black hole.}
\label{S2}
In this section, we will present the extension of the abelian Higgs model applied to acoustic metrics, with modification of the scalar field through Lorentz symmetry violation. As demonstrated in \cite{anacleto2010acoustic}, the Lagrangian for the Abelian Higgs model with Lorentz symmetry violation is given by:
\begin{eqnarray}
\mathcal{L} = -\dfrac{1}{4}F_{\mu\nu}F^{\mu\nu} + |D_{\mu}\phi|^{2} + m^{2}|\phi|^{2} - b|\phi|^{4} + k^{\mu\nu}D_{\mu}\phi^{*}D_{\nu}\phi,
\end{eqnarray}
where $F_{\mu\nu} = \partial_{\mu}A_{\nu} -\partial_{\nu}A_{\mu}$, $D_{\mu}\phi = \partial_{\mu}\phi - ieA_{\mu}\phi$ and $k^{\mu\nu}$ is a constant symmetric tensor that implements Lorentz symmetry breaking.
We reduce the ten components of the tensor $k_{\mu \nu}$ independent components, choosing the following entries: $k_{ii} = k_{00}\equiv \beta$ and $k_{0i}=k_{ij}\equiv \alpha$.
Following the procedure in by~\cite{Casana:2011bv}, to estimate upper bounds for Lorentz violation parameters in the context of relativistic and non-relativistic Bose-Einstein condensation, the components of $k_{\mu\nu}$ are expected to satisfy $tr(k_{ij}) \leq 3 \times 10^{-6}$.
However, our main objective is to investigate the qualitative effects of Lorentz symmetry violation on absorption and quasinormal modes. To make these effects numerically significant and clearly identifiable in our results, we will treat $\alpha$ and $\beta$ as free parameters. We will explore large illustrative values that allow us to amplify and better understand the impact of Lorentz violation on the system's dynamics.
The tensor is then given by 
\begin{eqnarray}
k_{\mu\nu} = \left[{\begin{array}{cccc}\beta & \alpha & \alpha & \alpha\\
\alpha & \beta & \alpha & \alpha\\
\alpha & \alpha & \beta & \alpha\\
\alpha & \alpha & \alpha & \beta\\
\end{array}}\right] \qquad (\mu, \nu = 0, 1, 2, 3),
\label{tensor}
\end{eqnarray}
where $\alpha$ and $\beta$ are real parameters.
Next, we will analyze the two-dimensional acoustic metric that describes a rotating acoustic black hole for the specific case where $\beta = 0$ and $\alpha \neq 0$. In this way, we describe the effects due to the presence of Lorentz symmetry breaking in the absorption cross sections and on quasinormal frequencies.

\subsection{Rotating acoustic black hole.}
In recent studies~\cite{campos2024absorption}, we studied the effect of Lorentz breaking on a metric for a canonical acoustic black hole, where we verified symmetry breaking in the absorption and differential scattering cross-section in three dimensions. In this work, we extend the analysis to the (2+1) dimension scenario. The fundamental metric for this context was initially proposed by~\cite{anacleto2010acoustic}. Subsequent works~\cite{anacleto2011superresonance, anacleto2012analogue, anacleto2019quantum} focused on the effects of superresonance and quantum corrections on planar acoustic metrics.
In this paper, we investigate the absorption cross section and quasinormal modes of a black hole with rotation in (2 + 1) dimensions. Assuming incompressibility and axial symmetry, such that the density $\rho$ is independent of position and the continuity equation implies a radial velocity dependence given by $v \propto 1/r$. Using the following conditions for the tensor $k_{\mu\nu}$ \eqref{tensor} $\beta = 0$ and $\alpha \neq 0$, and considering the non-relativistic limit $(v^{2} << c_{s}^{2})$, we obtain the following line element for acoustics with Lorentz symmetry breaking: \citep{anacleto2010acoustic} 

\begin{eqnarray}
ds^{2}=-\left(1 + \alpha\right)\left[c_{s}^{2} - \dfrac{(v_{r}^{2} + v_{\phi}^{2})}{(1+\alpha)}\right] d\tau^{2} - 2v_{\phi}rd\phi d\tau + \dfrac{c_{s}^{2} dr^{2}}{\left[c_{s}^{2} - \dfrac{v_{r}^{2}}{(1+\alpha)}\right]} + \left[1-2\alpha(v_{r} + v_{\phi})\right] r^{2}d\phi^{2}.
\label{ds1}
\end{eqnarray}
For the (2+1)-dimensional case, the fluid velocity is defined as $v^{2} = v_{r}^{2} + v_{\phi}^{2} = (A^{2} + B^{2})/r^{2}$, where $A$ and $B$ are parameters related to radial drainage and fluid vortex, respectively. For the following steps $c_{s}=1$, the line element \eqref{ds1} can be rewritten in the form
\begin{eqnarray}
ds^{2}= -F(r)(1+\alpha)d\tau^{2} - 2Bd\phi d\tau + G^{-1}(r)dr^{2} + \gamma^{2}(r)d\phi^{2},
\label{ds2}
\end{eqnarray}
where metric functions are defined as
\begin{eqnarray}
F(r) = 1 - \dfrac{(A^{2} + B^{2})}{(1+\alpha)r^{2}}, \qquad G(r) = 1 -\dfrac{A^{2}}{(1+\alpha)r^{2}}, \quad \text{and} \quad \gamma(r) = r\left[1 + \dfrac{2\alpha(A+B)}{r}\right]^{1/2}.
\end{eqnarray}
The line element \eqref{ds2} describes  a rotating acoustic black hole with Lorentz breakdown. The ergo-region is located at $r < r_{e}$, with radius given by $r_{e} = \sqrt{(A^{2} + B^{2})/(1+\alpha)}$, while the sonic event horizon is given by $r_{h} = A/\sqrt{1+\alpha}$.
An alternative definition for the horizon, in terms of the function $\gamma$ is
\begin{eqnarray}
    \gamma_{h} = (1+\alpha)^{-1/2} \sqrt{A^{2}+2\alpha\sqrt{1+\alpha} A(A+B)}.
    \label{gamah}
\end{eqnarray}
\section{Absorption }
\label{S3}
\begin{figure}[htbp]
 \centering
\includegraphics[width=.4\textwidth]{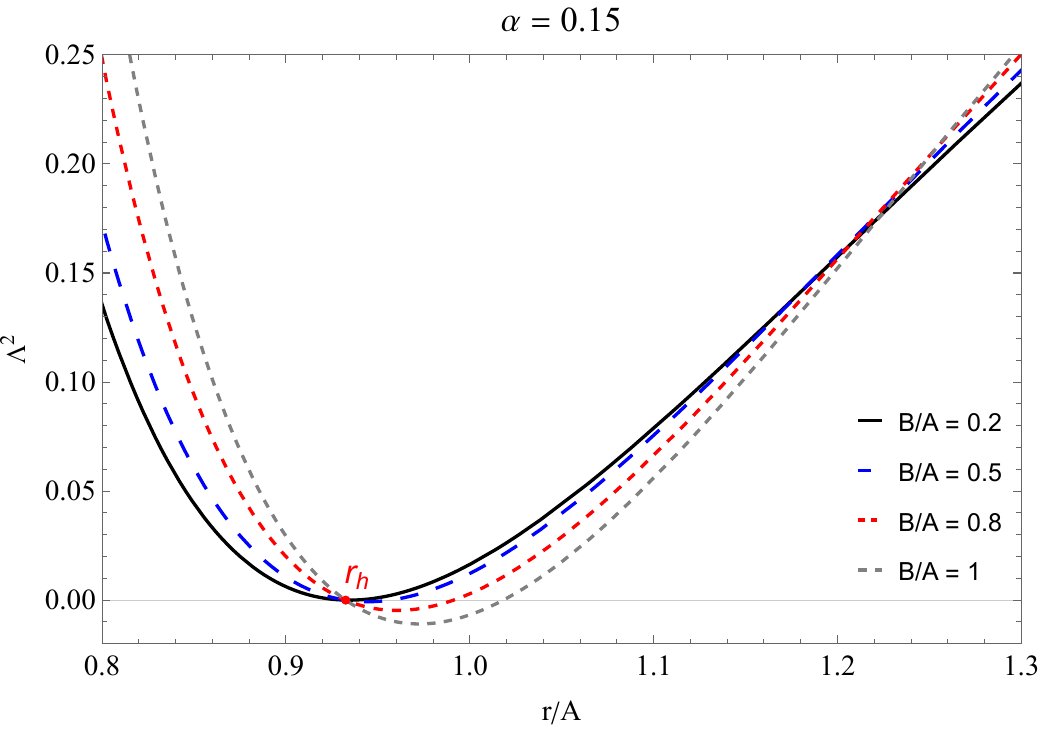}
   \caption{\footnotesize{Evolution of the function $\Lambda^{2}$ for a fixed value of the Lorentz violation parameter $\alpha=0.15$ and different values for the rotation $B$. The negative region of $\Lambda^2$ for $r>r_{h}$ shrinks as $B$ decreases.}}
 \label{plotLamb}
\end{figure}

In this section, we will analyze the absorption cross section analytically in the low-frequency limit and in the high-frequency regime using geodesic analysis. For a complete frequency regime, we will numerically analyze the absorption cross section, thus verifying the effects of Lorentz symmetry breaking. We begin by applying the partial wave method to a massless scalar field. Using the metric \eqref{ds2} in the Klein-Gordon equation $\frac{1}{\sqrt{-g}}\partial_{\mu}\left(\sqrt{-g}g^{\mu\nu}\partial_{\nu}\Psi \right)=0$ and proposing the following variable separation.

\begin{equation}
\Psi\left(\tau,r,\phi\right) = \frac{\psi_{\omega m}(r)}{\sqrt{\gamma(r)}}e^{i(m\phi-\omega \tau)},
\end{equation}
where $m = 0, \pm1,\pm2, \dots$ is the azimuthal quantum number, the radial equation is obtained for $\psi_{\omega m}$  of the form 
\begin{equation}
\Lambda(r)\frac{d}{dr}\left[\Lambda(r)\frac{d\psi_{\omega m}}{dr}\right] + \left\lbrace\left(\omega - \dfrac{m B}{\gamma^{2}(r)\sqrt{1+\alpha}}\right)^{2} - V(r)\right\rbrace\psi_{\omega m} = 0,
\label{eqRadCompleta}
\end{equation}
with $\Lambda = \sqrt{\left(\dfrac{B^{2}}{(1+\alpha)\gamma^{2}(r)}+F(r)\right)G(r)}$. The potential $V$ is given by
\begin{eqnarray}
    V=\dfrac{\Lambda(r)}{\gamma(r)}\left[\dfrac{m^{2}}{\gamma(r)} + \dfrac{\gamma'(r)\Lambda'(r)}{2}-\dfrac{\Lambda(r)}{4\gamma(r)}\left(\gamma'(r)^{2} - 2\gamma(r)\gamma''(r)\right) \right],
\end{eqnarray}
and the derivatives of $\gamma(r)$ are:
\begin{eqnarray}
\gamma'(r) = \dfrac{r+(A+B)\alpha}{\gamma(r)}, \qquad \gamma''(r) = \dfrac{-\alpha^{2}(A+B)^{2}}{\gamma^{3}(r)}.
\end{eqnarray}
The function $\Lambda^2$ exhibits a negative region near the horizon $r_{h}$, for values of $r>r_{h}$, as show in the Fig. \ref{plotLamb}. This makes the radial equation \eqref{eqRadCompleta} complex in this region. In the Fig. \ref{plotLamb}, we assume $\alpha=0.15$ and vary the rotation parameter $B$; the negative region is reduced in the slow rotation regime. In this way, we can circumvent this pathology in the radial equation by rewriting it assuming low rotations.
In the limit of slow rotations and small $\alpha$, the radial equation above can be simplified. Eliminating combinations of $\alpha B^2$, the expanded metric function becomes $\Lambda \approx 1 - A^{2}/(1+\alpha)r^{2}$. Similarly, expanding the potential, we obtain
\begin{equation}
\Lambda(r)\frac{d}{dr}\left[\Lambda(r)\frac{d\psi_{\omega m}}{dr}\right] + \left[\left(\omega - \dfrac{m B}{\sqrt{1+\alpha}\gamma^{2}(r)}\right)^{2} - V_{eff}\right]\psi_{\omega m} = 0,
\label{eqRad2}
\end{equation}
with the effective potential given by:
\begin{eqnarray}
V_{eff} = \dfrac{\Lambda(r)}{\gamma^{2}(r)}\left[m^{2} + \dfrac{\left(r+\alpha(A+B)\right)A^{2}}{(1+\alpha)r^{3}} - \dfrac{\Lambda(r)}{4} \right].
\label{poteff}
\end{eqnarray} 
Introducing a new coordinate $dx=\frac{dr}{\Lambda(r)}$ called tortoise $x = r + \dfrac{A}{2\sqrt{1+\alpha}}\log\Big|\dfrac{r\sqrt{1+\alpha} - A}{r\sqrt{1+\alpha} + A}\Big|$. The radial equation can be transformed into a Schr\"odinger-type equation of the form
\begin{eqnarray}
\dfrac{d^{2}\psi_{\omega m}}{dx^{2}} + \left[\left(\omega - \dfrac{m B}{\sqrt{1+\alpha}\gamma^{2}(r)}\right)^{2} - V_{eff}\right]\psi_{\omega m} = 0.
\label{eqRSchro}
\end{eqnarray}

\begin{figure}[htbp]
 \centering
\subfigure{\includegraphics[width=.45\textwidth]{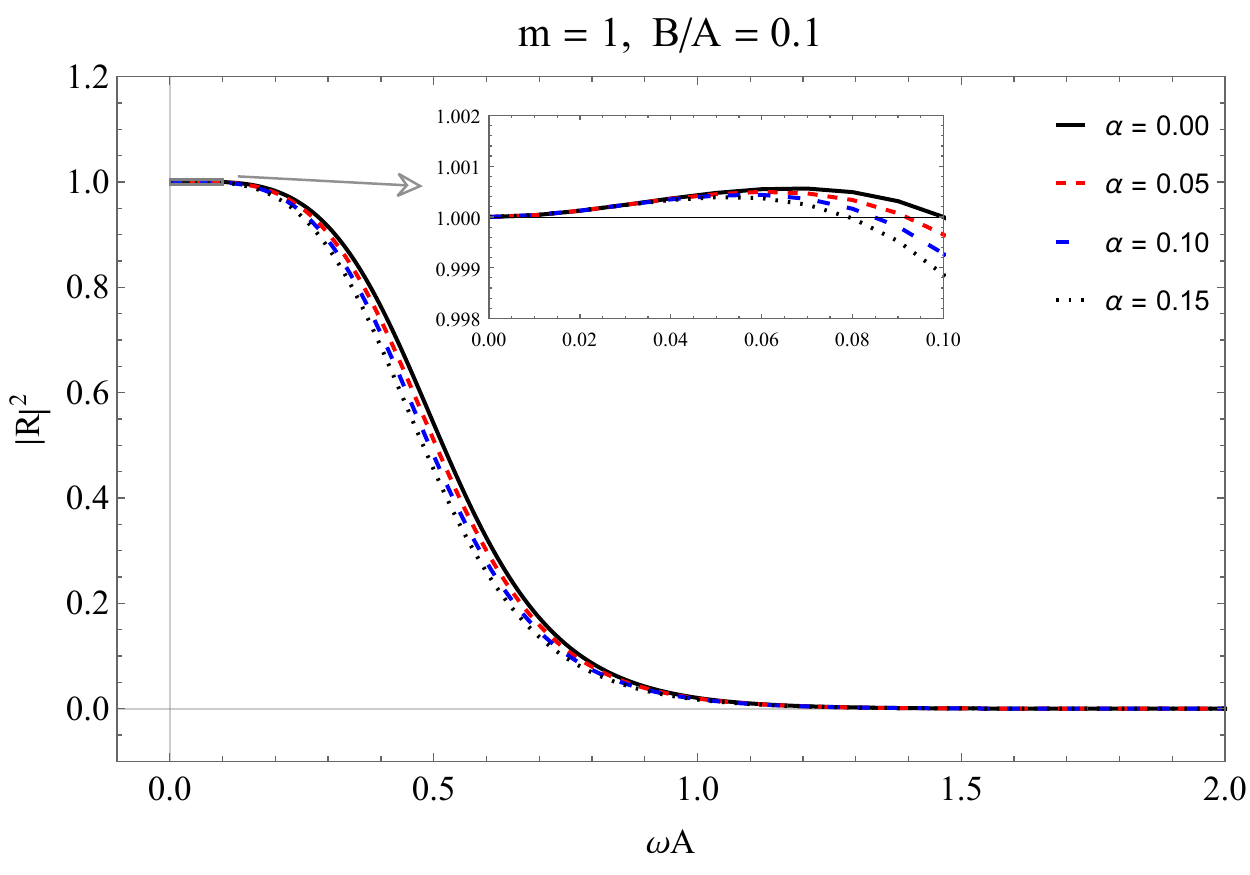}\label{Res_m1_B01}}
 \quad
 \subfigure{\includegraphics[width=.45\textwidth]{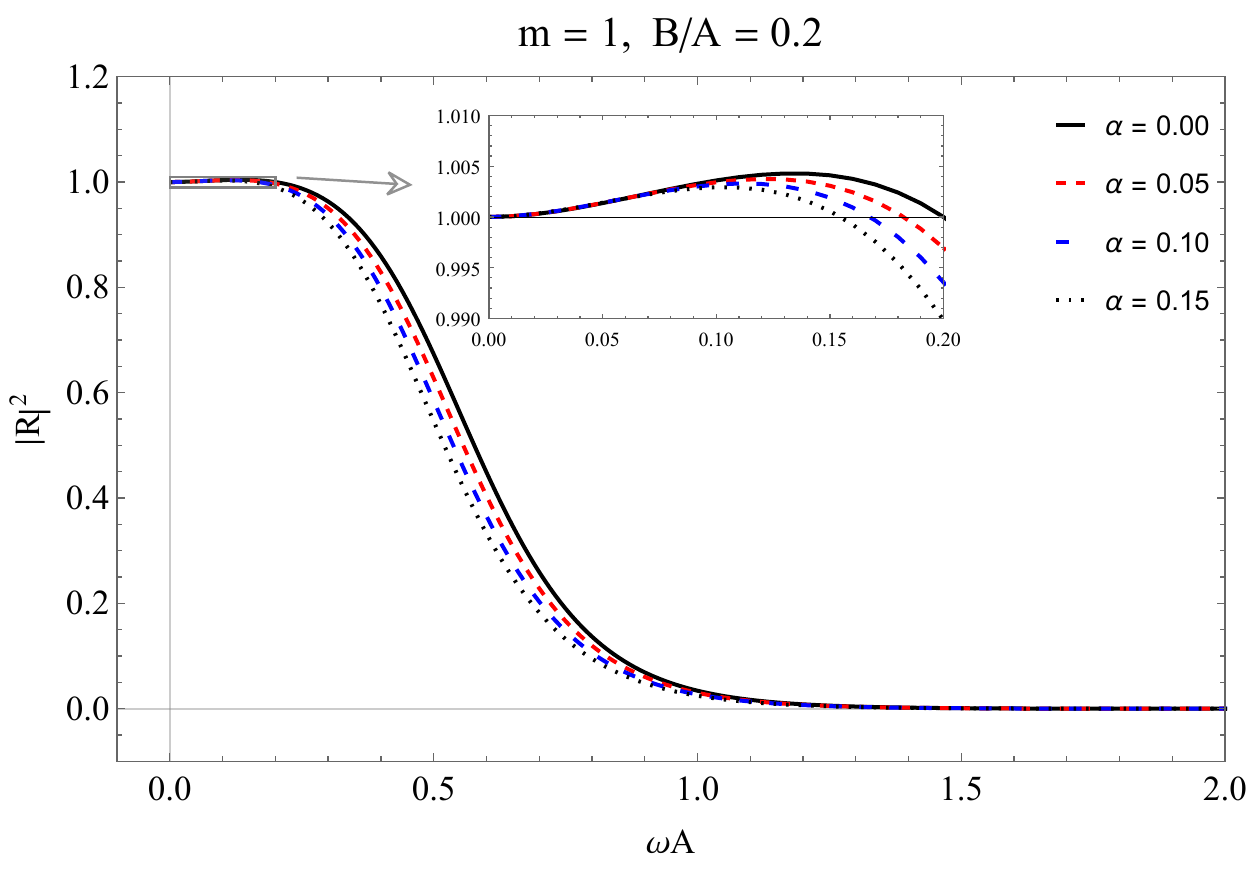}\label{Res_m1_B02}}
  \caption{\footnotesize{Reflection coefficient for a rotating acoustic black hole with Lorentz symmetry breaking. In both panels, the effect of the parameter $\alpha$ on superradiance is observed $|R_{\omega m}|^{2}>1$}.}
 \label{refl}
\end{figure}

Analyzing the asymptotic cases for the equation \eqref{eqRSchro}. Near the horizon $(r \rightarrow A/\sqrt{1+\alpha})$ the gamma function takes the form \eqref{gamah}, the effective potential $V_{eff} \rightarrow 0$. Under these conditions, the solution of the equation \eqref{eqRSchro} is given by the form
\begin{equation}
    \psi_{\omega m} \approx T_{\omega m} e^{-i\left(\omega -  \dfrac{mB}{\sqrt{1+\alpha}\gamma_{h}^{2}} \right)x},
    \label{solrh}
\end{equation}
where $|T_{\omega m}|^{2}$ is the transmission coefficient, related to the fraction of the wave that  crosses the horizon. The connection between the coefficients of transmission and reflection can be obtained as follows: 
\begin{eqnarray}
    |R_{\omega m}|^{2} = 1 - \left(1-\dfrac{m B}{\omega \sqrt{1+\alpha}\gamma_{h}^{2}} \right)|T_{\omega m}|^{2}.
\end{eqnarray}
We have a superradiance regime when $|R_{\omega m}|^{2}>1$, this occurs when the frequency $\omega$ is less than the critical frequency $\omega_{c}$~\cite{basak2003reflection}. This critical frequency for the superradiance regime is defined as $\omega_{c} = m \sqrt{1+\alpha}B/\gamma_h^2$ where $\gamma_h$ is the value of the function $\gamma(r)$ on the horizon. Figure \ref{refl} illustrates the behavior of the reflection coefficient as a function of frequency $\omega$. It can be observed that increasing the Lorentz breaking parameter $\alpha$ reduces the superradiance caused by rotation.

For the asymptotic limit at infinity $(r \rightarrow \infty)$ the solution for \eqref{eqRSchro} is given in the form
\begin{eqnarray}
    \psi_{\omega m} \approx e^{-i\omega x } + R_{\omega m}e^{i\omega x}.
    \label{solinf}
\end{eqnarray}
 With asymptotic solutions at both the horizon and spatial infinity, we can now determine the absorption cross section both analytically and numerically, as well as obtain the quasinormal modes. In the following section, we will analyze the absorption cross section in the low-frequency limit.
\subsection{Low frequency  absorption cross section}
\begin{figure}[htbp]
 \centering
 \includegraphics[width=.45\textwidth]{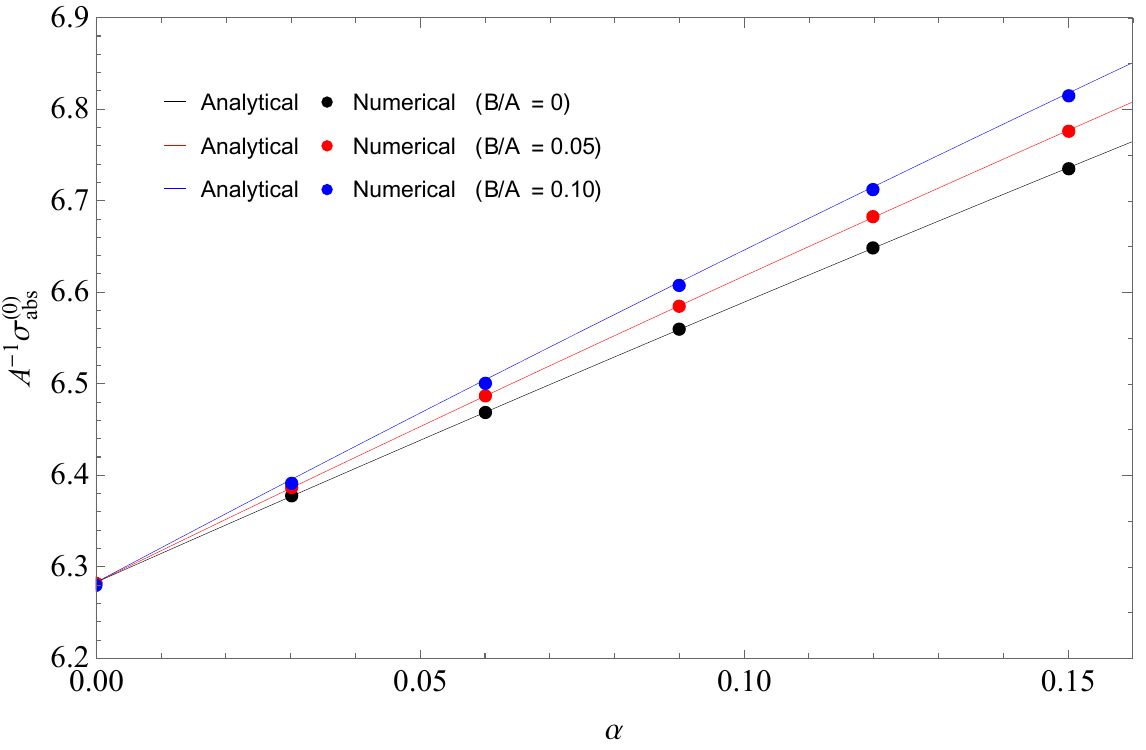}
 \caption{\footnotesize{Results for absorption cross-section for the low frequency regime. The lines represent the results of the analytical equation, varying the rotation parameter. The points are numerical results for this regime and the respective values of $B$.}}
 \label{absLF}
\end{figure}

To obtain the absorption cross-section in the low-frequency regime, we perform the following coordinate change $\gamma = \sqrt{r(r+2\alpha(A+B))} $ where $r \approx \gamma-(A+B)\alpha$. With this transformation, the radial equation \eqref{eqRad2} takes the form
\begin{eqnarray}
   &&\dfrac{\left(A^{2}-(1+\alpha)(\gamma-(A+B)\alpha)^{2}\right)^{2}}{(1+\alpha)^{2}\left(\gamma-(A+B)\alpha\right)^{4}} \dfrac{d^{2}\psi_{\omega m}}{d\gamma^{2}} - \dfrac{2A^{2}\left(A^{2}-(1+\alpha)(\gamma -(A+B))^{2}\right)}{(1+\alpha)^{2}\left(\gamma-(A+B)\alpha\right)^{5}}\dfrac{d\psi_{\omega m}}{d\gamma}\nonumber \\&\quad &+ \left[\left(\omega - \dfrac{Bm}{\sqrt{1+\alpha}\gamma^{2}}\right)^{2}-V_{eff}(\gamma)\right]\psi_{\omega m} = 0.
\end{eqnarray}
Let's determine the asymptotic behavior of $\psi_{\omega m}$. In the limit $\gamma \rightarrow \infty$, the equation above reduces to
\begin{eqnarray}
    \dfrac{d^{2}\psi_{\omega m}}{d\gamma^{2}} + \left(\omega^{2} + \dfrac{4-4m^{2}-8B m \omega/\sqrt{1+\alpha}}{4\gamma^{2}} \right)\psi_{\omega m} = 0.
\end{eqnarray}
The general solution to this equation can be expressed in terms of the Hankel function~\cite{arfken2005mathematical}.
\begin{eqnarray}
    \psi_{\omega m} \approx \sqrt{\dfrac{\pi\omega\gamma}{2}}\left[e^{-i(\nu+1/2)\pi/2}H_{\nu}^{(1)*}(\omega \gamma)+e^{i(\nu+1/2)\pi/2}R_{\omega m}H_{\nu}^{(1)}(\omega \gamma)\right],
    \label{solpsi}
\end{eqnarray}
where $\nu =|\sqrt{m(m+2B\omega)}|$ and $H_{\nu}^{(1)}(\omega \gamma)$ is the Hankel function of the first kind.
For the low frequency regime, we consider $m=0$. In the limit $\omega\gamma <<1$ the Hankel function behaves as $H_{\nu}^{(1)}(\omega \gamma)\approx 1+ (2i/\pi)\left[\xi +\log(\omega \gamma/2)\right]$, where $\xi$ is a constant~\cite{arfken2005mathematical}. Substituting this expansion in \eqref{solpsi}, we obtain
\begin{eqnarray}
\psi_{\omega m} \approx \dfrac{\sqrt{\pi \omega \gamma}}{2}\left[(1-i)+(1+i)R_{\omega 0} + \dfrac{2i}{\pi}\left[\xi+\log(\omega \gamma/2)\right]\left[(1+i)R_{\omega 0} - (1-i)\right]\right].
\end{eqnarray}
For $\psi_{\omega 0}$ to be finite in the low-frequency limit, it is necessary that $R_{\omega 0}\approx-i + \dots + \mathcal{O}(\omega)$. With this condition, the solution simplifies to
\begin{eqnarray}
    \psi_{\omega 0} \approx \sqrt{\pi \omega \gamma}(1-i).
    \label{sol_lf}
\end{eqnarray}
Comparing the solution \eqref{sol_lf} with the asymptotic solution \eqref{solrh}, applying $m=0$ and the limit $\omega x<<1$ such that $e^{-i\omega x}\approx 1 + \mathcal{O}(\omega)$. Furthermore, we consider $r \rightarrow A/\sqrt{1+\alpha}$ in the equation \eqref{sol_lf} such that $\gamma = (1+\alpha)^{-1/2} \sqrt{A^{2}+2\alpha\sqrt{1+\alpha} A(A+B)}$. Comparing the two equations, we obtain the transmission coefficient for this regime
\begin{eqnarray}
    T_{\omega 0} \approx (1-i)\sqrt{\pi\omega(1+\alpha)^{-1/2} \sqrt{A^{2}+2\alpha\sqrt{1+\alpha} A(A+B)}}.
\end{eqnarray}
Using the relation for the partial wave absorption cross section, $\sigma^{m}_{abs} = |T_{\omega m}|^{2}/\omega$,
and substituting the transmission coefficient found, we obtain the absorption cross section for low frequencies.
\begin{eqnarray}
    \sigma_{abs}^{0} = \dfrac{|T_{\omega 0}|^{2}}{\omega} = \dfrac{2\pi A}{\sqrt{1+\alpha}}\sqrt{1+\dfrac{2\alpha\sqrt{1+\alpha}(A+B)}{A}}.
    \label{analLF}
\end{eqnarray}
We have therefore verified that the Lorentz breakdown influences the absorption cross-section in the low-frequency regime, introducing an explicit dependence on the rotation parameter $B$. In figure \ref{absLF} the analytical results obtained from equation \eqref{analLF} are compared with the numerical results. We observe that absorption increases with the parameter $\alpha$ and that the rotation $B$ contributes to the increase in absorption even at low frequencies. 
\subsection{Null geodesic analysis}
We can obtain classical scattering at high energies by studying geodesic scattering.
The equations of motion for a particle in a (2+1)-dimensional acoustic metric were initially analyzed in \cite{dolan2009scattering,oliveira2010absorption}. 
In this section, we investigate the influence of the Lorentz breaking parameter $\alpha$ on the geodesic trajectories of the modified (2+1) acoustic metric draining bathtub (DBT), by numerically solving the orbital equations.

Geodesics are obtained from the Lagrangian $\mathcal{L} \equiv \dfrac{1}{2}g_{\mu\nu}\dot{x}^{\mu}\dot{x}^{\nu}$ associated with the metric \eqref{ds2}, resulting in the following form.
\begin{equation}
2\mathcal{L} = -F(r) (1+\alpha)\dot{\tau}^{2} - 2B\dot{\phi} \dot{\tau} + G^{-1}(r)\dot{r}^{2} + \gamma^{2}(r)\dot{\phi}^{2},
\label{elidot}
\end{equation} 
where ``." is the derivative with respect to the affine parameter.
For a null geodesic (sound ray) moving in an equatorial plane $\theta = \pi/2$, two motion constants, associated with the symmetries of the metric, can be identified:
\begin{equation}
E = \sqrt{1+\alpha}\left(1-\dfrac{(A^{2}+B^{2})}{(1+\alpha)r^{2}}\right)\dot{\tau} + B\dot{\phi}, \qquad  L = -B\dot{\tau} +  \gamma^{2}(r)\dot{\phi}.
\label{EL}
\end{equation}
For a null geodesic where $g_{\mu\nu}\dot{x}^{\mu}\dot{x}^{\nu} = 0$, and using the equations \eqref{EL} we obtain the following relations
\begin{eqnarray}
\dot{\tau} =\dfrac{-r^{2}\left(BL - \sqrt{1+\alpha}E\gamma^{2}(r)\right)}{((1+\alpha)r^{2} - A^{2})\gamma^{2}(r) + B^{2}(r^{2}-\gamma^{2}(r))} , \qquad \dot{\phi} = -\dfrac{(A^{2}+ B^{2}-(1+\alpha)r^{2})L - \sqrt{1+\alpha}BEr^{2}}{((1+\alpha)r^{2}-A^{2})\gamma^{2}(r)+B^{2}(r^{2}-\gamma^{2}(r))} \quad \text{and}
\label{eq_t_phi}
\end{eqnarray}
\begin{eqnarray}
\dot{r}^{2}  = \dfrac{((1+\alpha)r^{2}-A^{2})\left[(A^{2}+B^{2} - (1+\alpha)r^{2})L^{2}- 2\sqrt{1+\alpha}B E L r^{2} + (1+\alpha)E^{2}\gamma^{2}r^{2}\right]}{r^{2}(1+\alpha)\left[((1+\alpha)r^{2} - A^{2})\gamma^{2}(r) + B^{2}(r^{2}-\gamma^{2}(r))\right]}. 
\label{eqdotR}
\end{eqnarray}
Using again the condition of low rotation $B<<1$ and small $\alpha$, we make the following approximation.

\begin{eqnarray}
 B^{2}(r^{2}-\gamma^{2}(r)) = B^{2}r^{2} - B^{2}r^{2} -2B^{2}\alpha(A+B)r \approx 0.   
\end{eqnarray}
Thus we organize the equation \eqref{eqdotR} in the form
\begin{eqnarray}
\dot{r}^{2} + V(r) = E^{2},
\end{eqnarray}
where $V(r) = \left[1 - \dfrac{A^{2}+ B^{2}}{(1+\alpha)r^{2}} \right]\dfrac{L^{2}}{\gamma^{2}(r)} + \dfrac{2BE L}{\sqrt{1+\alpha}\gamma^{2}(r)}$.
Considering a null geodesic originating from infinity, we define the corresponding impact parameter $b$. In the zero-rotation limit ($B=0$), the impact parameter is simply $b = L/E$. For the rotating case, it is convenient to adopt the definition $b\equiv L/E + B$, so that, for geodesics with zero orbital angular momentum ($L=0$), the impact parameter is proportional to the rotation of the black hole, $b = B$. In this way, in the context of rotating acoustic black holes, they play the same role as radial geodesics in the static case. Geodesics with $b>B$ are co-rotating, while those with $b<B$ are those that counter-rotate with the acoustic black hole.
We have three interesting cases for scattering: if $b$ is large, the geodesic will be scattered; If $b$ is small, the geodesic will be absorbed; in the intermediate regime, the geodesic is in a critical orbit with radius $r = r_{c}$. For the critical case, we have the following conditions:
\begin{eqnarray}
V(r_{c}) = E^{2}\qquad \text{and} \qquad \dfrac{dV(r)}{dr}\Big|_{r=r_{c}} = 0.
\end{eqnarray}
Using the result found for the potential $V$, we obtain the critical impact parameter and the critical radius.
\begin{eqnarray}
b_{c}^{\pm} \approx \dfrac{-B \pm 2\sqrt{A^2 + B^2}}{\sqrt{1+\alpha}} - \dfrac{\alpha(A+B)\sqrt{2A^2+2B\left(B \mp \sqrt{A^2 + B^2}\right)}}{B \mp \sqrt{A^2 + B^2}},
\label{b_rcritico}
\end{eqnarray}
\begin{eqnarray}
r_{c}^{\pm} \approx \sqrt{\dfrac{2A^2+2B\left(-B \pm \sqrt{A^2 + B^2}\right)}{(1+\alpha)}} - \dfrac{\alpha(A+B)\left(B \mp \sqrt{A^2 + B^2}\right)}{B \pm \sqrt{A^2 + B^2}}.
\label{c_rcritico}
\end{eqnarray}

\begin{figure}[htbp]
 \centering
\subfigure{\includegraphics[width=.2\textwidth]{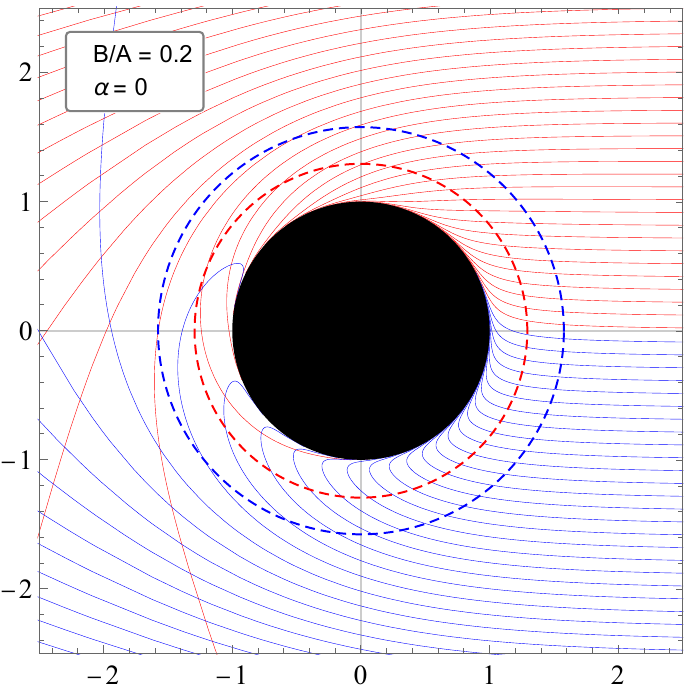}\label{geoB02a0}}
 \subfigure{\includegraphics[width=.2\textwidth]{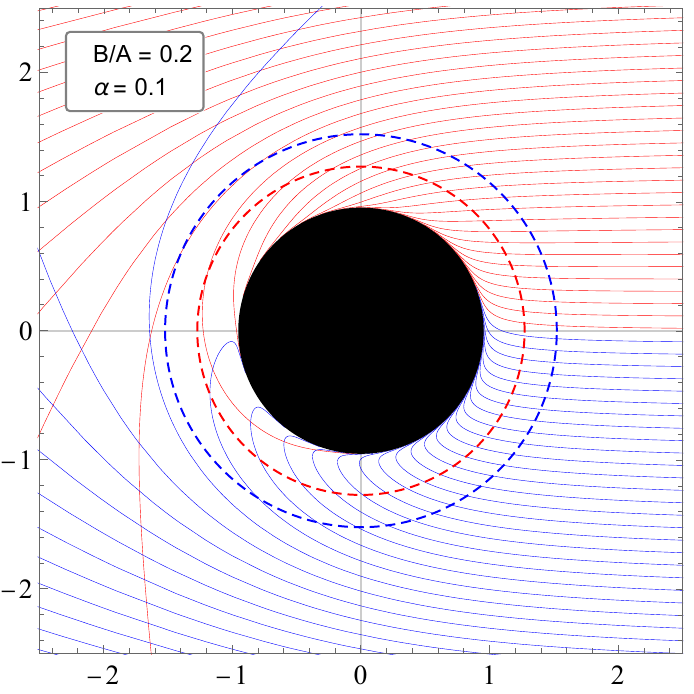}\label{geoB02a01}}
 \subfigure{\includegraphics[width=.2\textwidth]{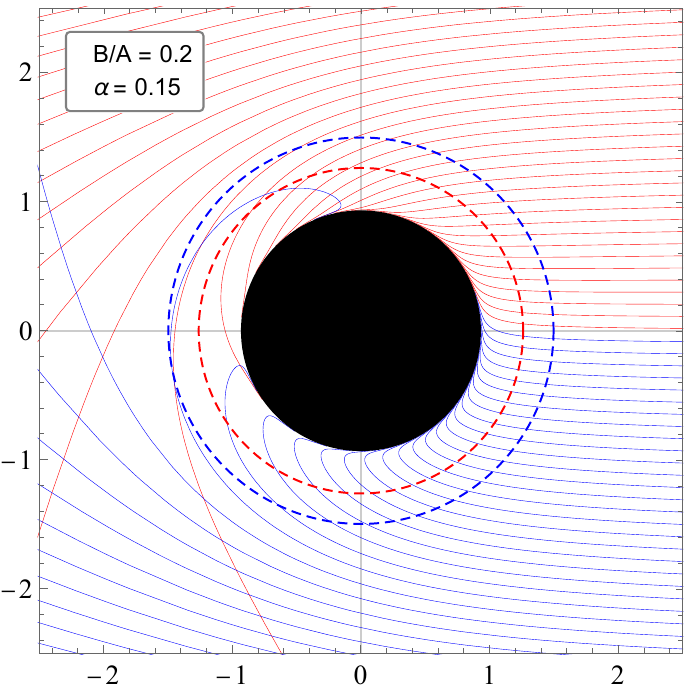}\label{geoB02a015}}
  \caption{\footnotesize{Geodesic lines for a (2+1)-dimensional acoustic black hole with Lorentz symmetry breaking. The blue lines represent the counter-rotating beams, while the red beams are co-rotating, just as the dotted circles represent the critical radii for each scenario. In the panels from left to right, we see the effect of the Lorentz breaking.}}
  \label{geodesic}
\end{figure}

The absorption section at high frequencies can be obtained using the co-rotating and counter-rotating critical impact parameter.
\begin{eqnarray}
\sigma_{abs}^{hf} = |b_{c}^{+}|+|b_{c}^{-}| \approx  4\sqrt{\dfrac{A^2 + B^2}{(1+\alpha)}}+\alpha\sqrt{2}(A + B)\left[\dfrac{\sqrt{A^2 + B\left(B + \sqrt{A^2 + B^2}\right)}}{B + \sqrt{A^2 + B^2}} - \dfrac{\sqrt{A^2 + B\left(B - \sqrt{A^2 + B^2}\right)}}{B - \sqrt{A^2 + B^2}}\right].
\label{asbhf}
\end{eqnarray}
Note that the critical impact parameter is modified by the Lorentz break so that the absorption cross section at high energies is also modified. As we saw in the low frequency regime, here the parameter corresponding to the Lorentz break increases the absorption cross-section. Since $\alpha =0$, we return to the usual case~\cite{oliveira2010absorption}.
By numerically solving equations \eqref{eq_t_phi} and \eqref{eqdotR} in the slow rotation regime, we obtain the geodesic trajectories shown in Figure \ref{geodesic}. We can verify the combined effect of rotation and Lorentz break on the behavior of geodesic curves, increasing the effective capture radius, analogous to the shadows of the black hole. We verify that the Lorentz break also influences the absorption cross section at high energies, tending to increase it, as will be verified numerically in the following section.

	\subsection{Numerical results}
We completed the analysis of the absorption cross section numerically for a full frequency range. For this, the absorption cross section is begin given by 
\begin{eqnarray}
    \sigma_{abs} = \dfrac{1}{\omega}\sum^{\infty}_{m=-\infty}\left(1-|e^{2i\delta_{m}}|^2\right).
    \label{absSeries}
\end{eqnarray}
We numerically obtain the phase shift
\begin{eqnarray}
e^{2i\delta_{m}}=i(-1)^{m}\left(\dfrac{C^{(out)}}{C^{(in)}}\right),
\end{eqnarray}
where the input coefficients $C^{(in)}$ and output coefficients $C^{(out)}$ come from the asymptotic solution \eqref{solinf} written in the form $\psi_{\omega m} \approx C^{(in)}e^{-i\omega x } + C^{(out)}e^{i\omega x}$.
The method begins with a series expansion at the horizon:
\begin{eqnarray}
    \psi(r)\approx e^{-i\left(\omega-mB/(\sqrt{1+\alpha})\gamma_{h}^{2}\right)x} \sum^{\infty}_{k=0} a_{k}(r-r_{h})^{k},
\end{eqnarray}
where $a_{k}$ are coefficients that can be determined analytically. We use this as an initial condition and integrate the radial equation \eqref{eqRad2} using the fourth-order Runge-Kutta method.
We compare the numerical result with the following analytical expansions:
\begin{eqnarray}
    &\psi^{(in)}(r)& \approx e^{-i\omega x} \sum^{\infty}_{k=0} \dfrac{z_{k}}{r^{k}},\\
&\psi^{(out)}(r)& \approx  e^{i\omega x}\sum^{\infty}_{k=0} \dfrac{z_{k}^{*}}{r^{k}},
\end{eqnarray}
where the coefficients $z_{k}$ and $z_{k}^{*}$ can be obtained analytically.
Setting up the following system in the regime of large $r$
\begin{equation}
    \begin{cases}
    C^{(in)}\psi^{(in)}(r_{inf}) + C^{(out)}\psi^{(out)}(r_{inf}) = \psi(r_{inf}),\\
    C^{(in)}{\psi^{(in)}}'(r_{inf}) + C^{(out)}{\psi^{(out)}}'(r_{inf}) = \psi' (r_{inf}),
    \end{cases}
\end{equation}
here $\psi (r_{inf})$ and its derivative $\psi' (r_{inf})$ result from numerical integration for large $r$ ($r_{inf}$), where we assume $r_{inf} = 200$. With the values of $C^{(in)}$ and $C^{(out)}$ obtained, we find the phase shift and consequently the absorption cross section.

\begin{figure}[htbp]
 \centering
\subfigure{\includegraphics[width=.45\textwidth]{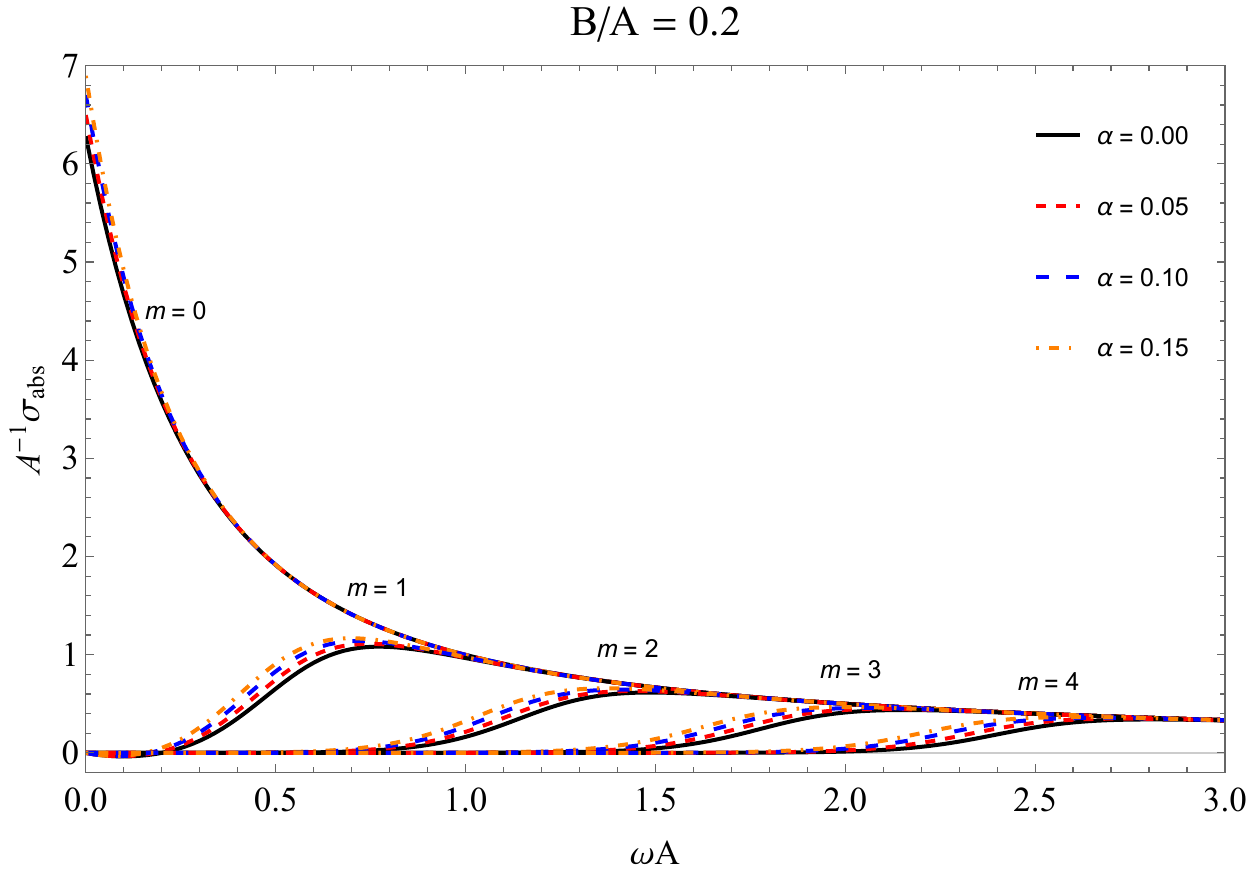}\label{Abs_B02_mp}}
 \quad
 \subfigure{\includegraphics[width=.45\textwidth]{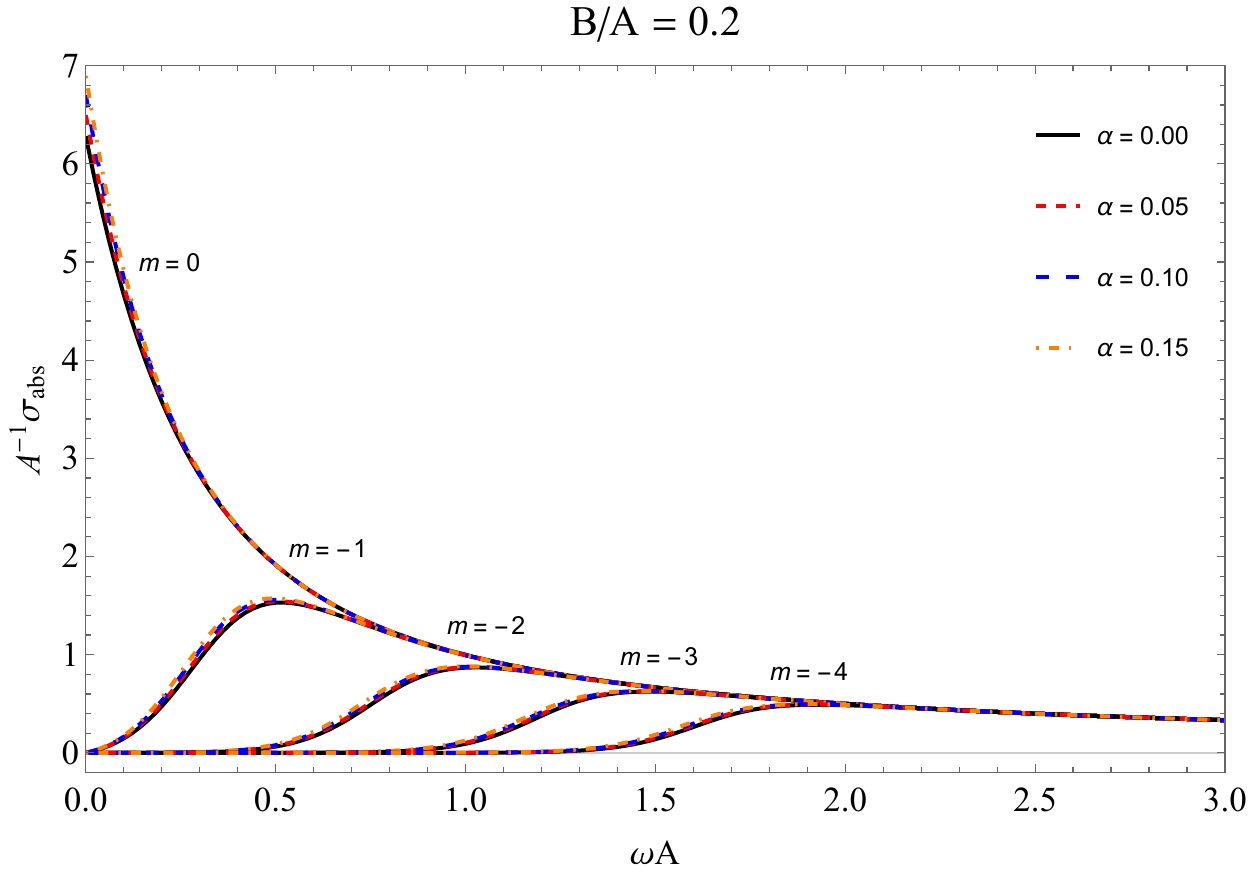}\label{Abs_B02_mm}}
  \caption{\footnotesize{Effect of Lorentz symmetry breaking in the partial absorption cross section, for some positive and negative values of $m$, assuming a small rotation $B=0.2$.}}
 \label{Abs}
\end{figure}

This method allows obtaining the reflection coefficient $R_{\omega m}$ and, consequently, the absorption cross section  for the entire frequency spectrum. 
Figure \ref{Abs} shows the partial absorption cross section as a function of frequency $\omega$, for the first modes with $m = 0, \pm 1, \pm2, \pm 3$. The parameters used were $B/A = 0.2$, illustrating the slow rotation regime. It can be observed that increasing the Lorentz breaking parameter $\alpha$ amplifies the absorption for both modes, but more pronouncedly for the co-rotating modes ($m > 0$).
This asymmetry between positive and negative $m$ is a signature of the acoustic black hole's rotation and is directly related to the superradiance phenomenon, discussed in Section \ref{S3}.

\begin{figure}[htbp]
 \centering
\subfigure{\includegraphics[width=.45\textwidth]{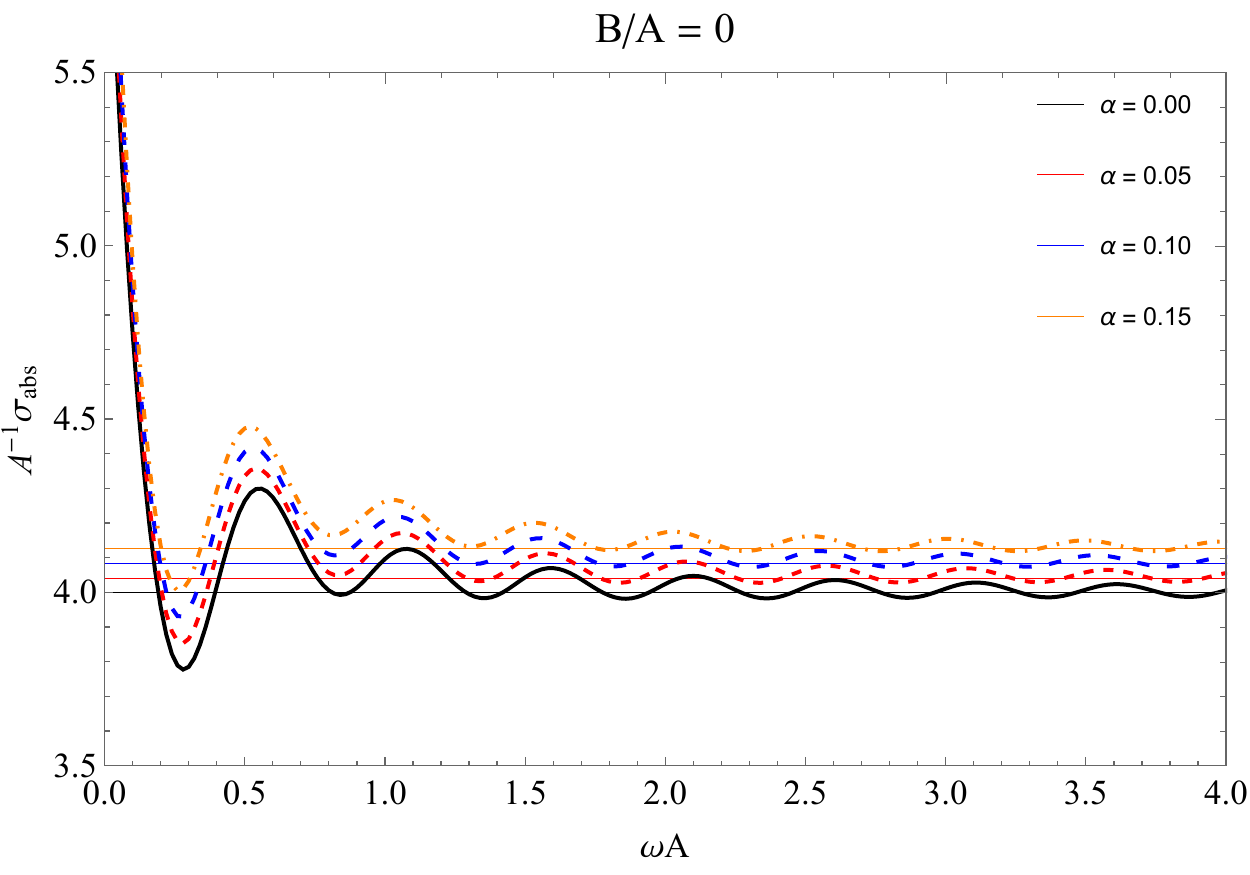}\label{AbsTotal_B0}}
 \quad
 \subfigure{\includegraphics[width=.45\textwidth]{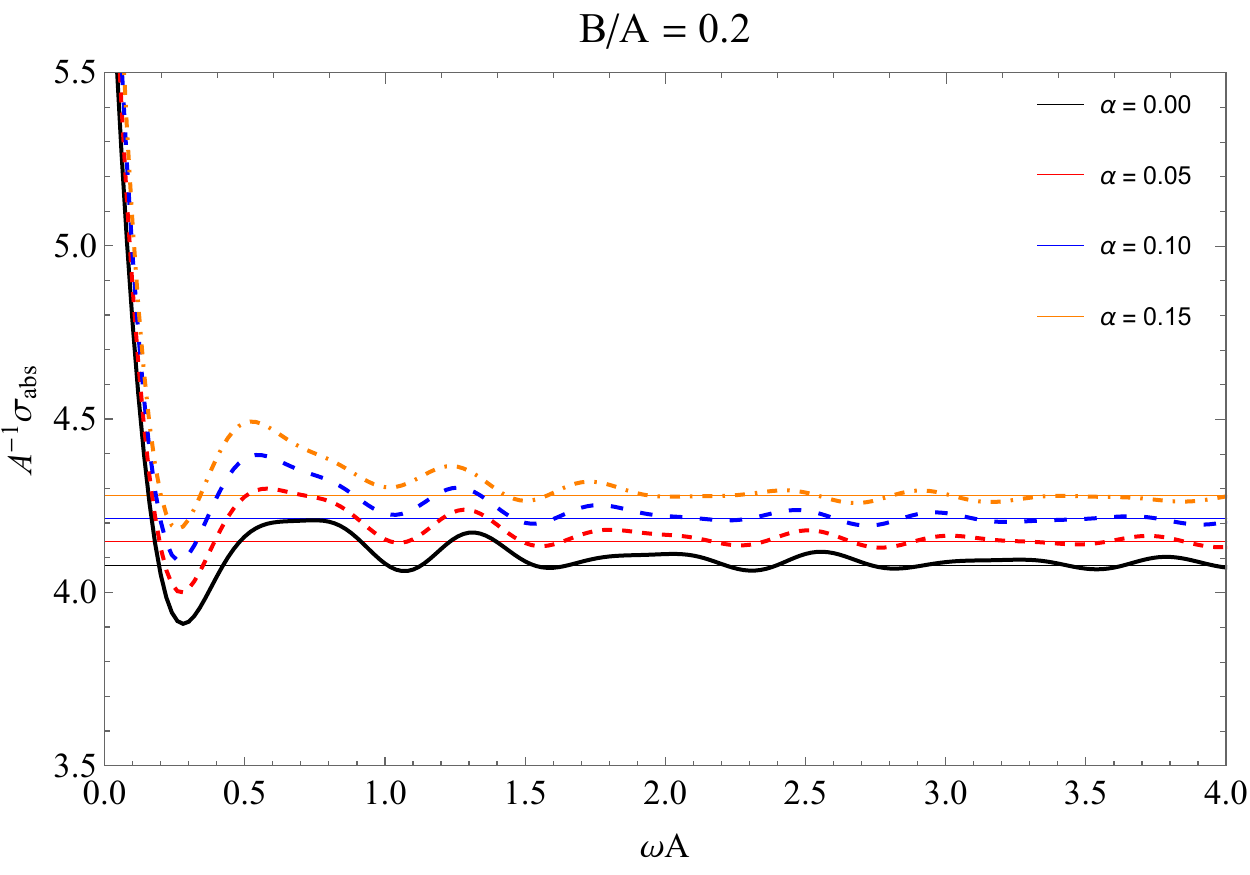}\label{AbsTotal_B02}}
  \caption{\footnotesize{Total absorption cross section for the case with and without rotation. The results are compared with the result for the absorption cross section at high energies (horizontal lines).}}
 \label{AbsTotal}
\end{figure}

The total absorption cross section, obtained by summing all partial modes, can be truncated by truncating the series \eqref{absSeries} at $m=-14$ to $m=14$. Figure \ref{AbsTotal} shows the numerical result for different values of $\alpha$, both in the absence ($B=0$) and in the presence ($B/A=0.2$) of slow rotation. In agreement with the analyses of the low and high frequency regimes, the numerical results confirm that the presence of the Lorentz breaking term ($\alpha \neq 0$) intensifies the absorption of waves by the acoustic black hole across the entire spectrum.
A consistency test is the verification of the high-frequency limit. As demonstrated in Figure \ref{AbsTotal}, as $\omega$ increases, the total absorption cross-section converges to the value predicted by the geodesic (classical) approximation of Eq. \eqref{asbhf}, represented by the horizontal lines. This convergence validates both the numerical procedure employed and the consistency between the wave (low frequencies) and geometric (high frequencies) analyses developed previously.

\section{Quasinormal modes for the rotating acoustic black hole}
\label{S4}
Quasinormal modes are solutions to the perturbation equations that satisfy specific boundary conditions: purely incoming waves at the event horizon and purely emergent waves at spatial infinity~\cite{berti2009quasinormal}. In terms of the turtle coordinate $x$, defined in Section \ref{S3}, these conditions are expressed as
\begin{eqnarray}
  \mathcal{\psi}_{\omega m}  \sim e^{-i\left(\omega-mB/(\sqrt{1+\alpha})\gamma_{h}^{2}\right)x} , \qquad (x \rightarrow  - \infty) \qquad \text{and} \qquad\mathcal{\psi}_{\omega m}  \sim e^{+i\omega x}, \qquad (x \rightarrow  + \infty),
\label{condQNM}
\end{eqnarray}
where the first condition corresponds to purely incoming waves at the horizon ($x \rightarrow - \infty$), while the second corresponds to purely outgoing waves at spatial infinity ($x \rightarrow +\infty$).
The quasinormal frequencies $\omega_{n}$ that satisfy these conditions form a discrete spectrum, indexed by the number of overtones $n = 0, 1, 2, \dots$. These frequencies are complex, with the real part, Re($\omega$), describing the oscillation frequency, and the imaginary part, Im($\omega$), describing the damping ratio of the mode. A study on the correspondence between null geodesics and quasinormal modes for the rotating acoustic black hole in 2+1 dimensions was demonstrated by~\cite{Dolan:2011ti}.

\subsection{WKB approximation}
\begin{figure}[htbp]
 \centering
\subfigure{\includegraphics[width=.45\textwidth]{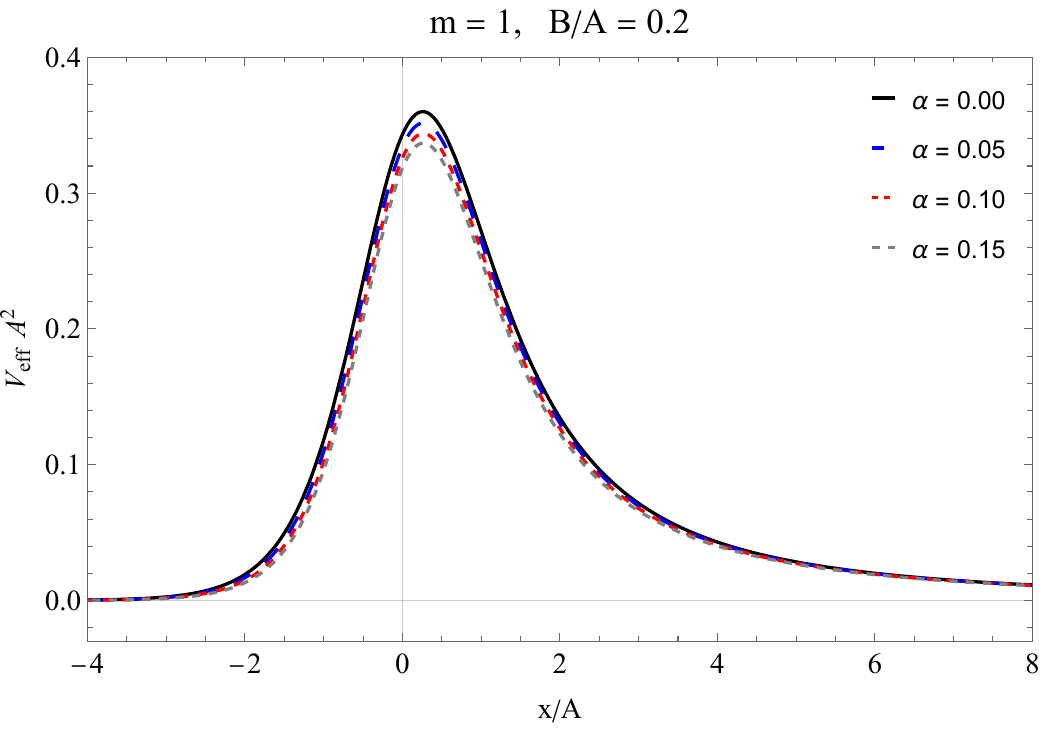}\label{plot_m1B02}}
 \quad
 \subfigure{\includegraphics[width=.45\textwidth]{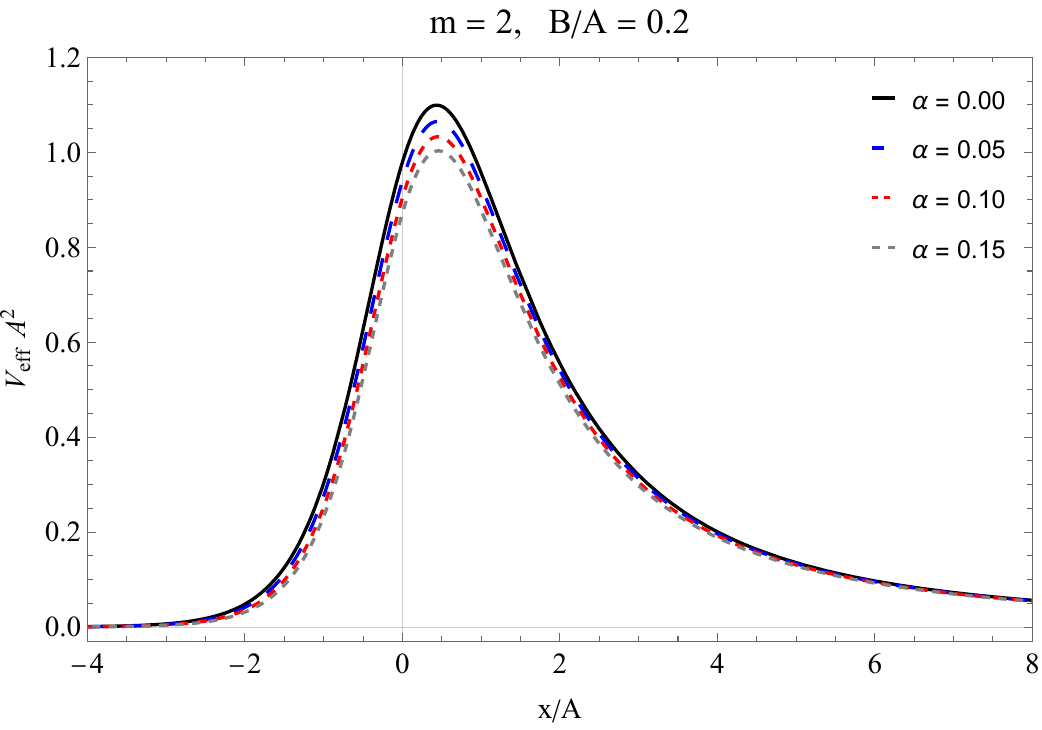}\label{plot_m2B02}}
  \caption{\footnotesize{The effective potential $V_{eff}$ as a function of the tortoise coordinate $x$.}}
 \label{potential}
\end{figure}

For the calculation of the quasinormal spectrum, we will use the WKB approximation, a technique widely used in the literature for providing satisfactory results for effective potentials that present a single barrier, as is the case of the potential $V_{eff}$ in \eqref{poteff}. Fig. \ref{potential} shows the behavior of the effective potential in terms of the tortoise coordinate $x$ for $B/A=0.2$, $m=1, 2$ and some values of the Lorentz violating parameter $\alpha$. A review of the higher-order WKB method can be found in~\citep{Konoplya:2019hlu, Konoplya:2026fqh}. We will use the corrected sixth-order approximation introduced by Konoplya~\citep{konoplya2003quasinormal}. However, we write the radial equation \eqref{eqRSchro} in the form
\begin{eqnarray}
\dfrac{d^{2}\psi_{\omega m}}{dx^{2}} + Q \psi_{\omega m} = 0,
\label{eqRSchro2}
\end{eqnarray}
where the generalized potential is $Q = \left(\omega - \dfrac{m B}{\sqrt{1+\alpha}\gamma^{2}(r)}\right)^{2} - V_{eff}(r)$. The WKB formula is written in the form
\begin{eqnarray}
\dfrac{iQ_{0}}{\sqrt{2Q''_{0}}} -\sum_{j=2}^{6} \Omega_{j} = n + \dfrac{1}{2},
\label{WKBformula}
\end{eqnarray}
where $Q_{0} \equiv Q(x_{0},\omega)$ is the value of the generalized potential at its maximum point, in tortoise coordinate $x$, $Q'' \equiv \left. d^2Q/dx^2\right|_{x=x_{0}}$ evaluated at the same point, and $\Omega_{j}$ are the higher-order correction terms.

Equation \eqref{WKBformula} is solved numerically using an iterative procedure. We use the value obtained from the WKB approximation for $B=0$, calculated analytically from the effective potential $V_{eff}$ as the initial guess for the frequency $\omega$. For each trial frequency, the maximum point $r_{0}$($x_{0}$) of the generalized potential is found by numerically solving the equation $Q'(r_{0},\omega)=0$ for the radial coordinate $r$. To calculate the derivative with respect to the tortoise coordinate $x$, we use the relation $d/dx=\Lambda(r)d/dr$. At the maximum point, the derivatives of $Q$ are calculated up to the order required for the sixth-order approximation, as are the correction terms $\Omega_{j}$. The complex frequency is obtained by solving equation \eqref{WKBformula} using the secant method. The iterative process is repeated until the relative variation between two consecutive iterations is less than $10^{-6}$, ensuring the stability and precision of the solution. The method is validated by the convergence of the results as the order of the WKB approximation increases (from 2nd to 6th order).
As discussed in~\cite{berti2004quasinormal}, the WKB method can still be used, producing good results, provided the rotation parameter is small. Since we are studying the slow rotation regime and small Lorentz violating parameter $\alpha$, our study falls within the limitations of the method.

\begin{figure}[htbp]
 \centering
\subfigure[]{\includegraphics[width=.40\textwidth]{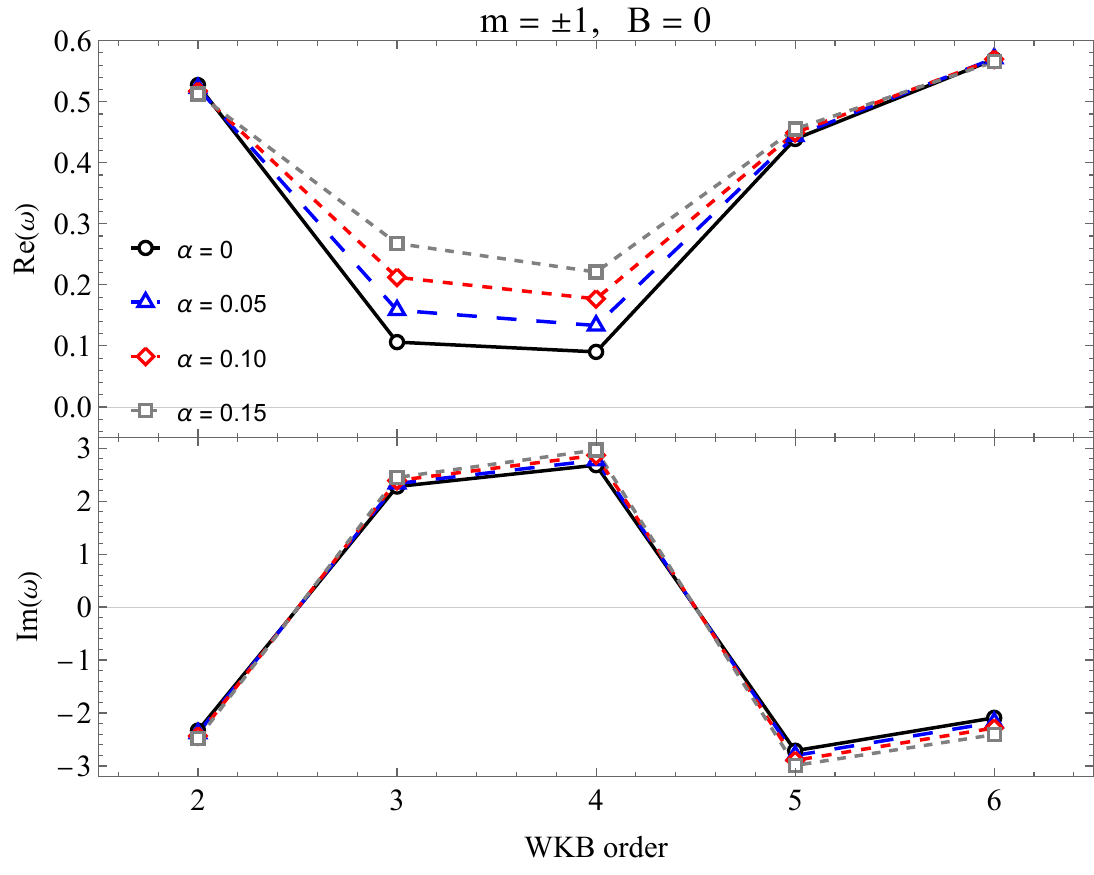}\label{WKBorder_m1B0n2}}
\quad
\subfigure[]{\includegraphics[width=.40\textwidth]{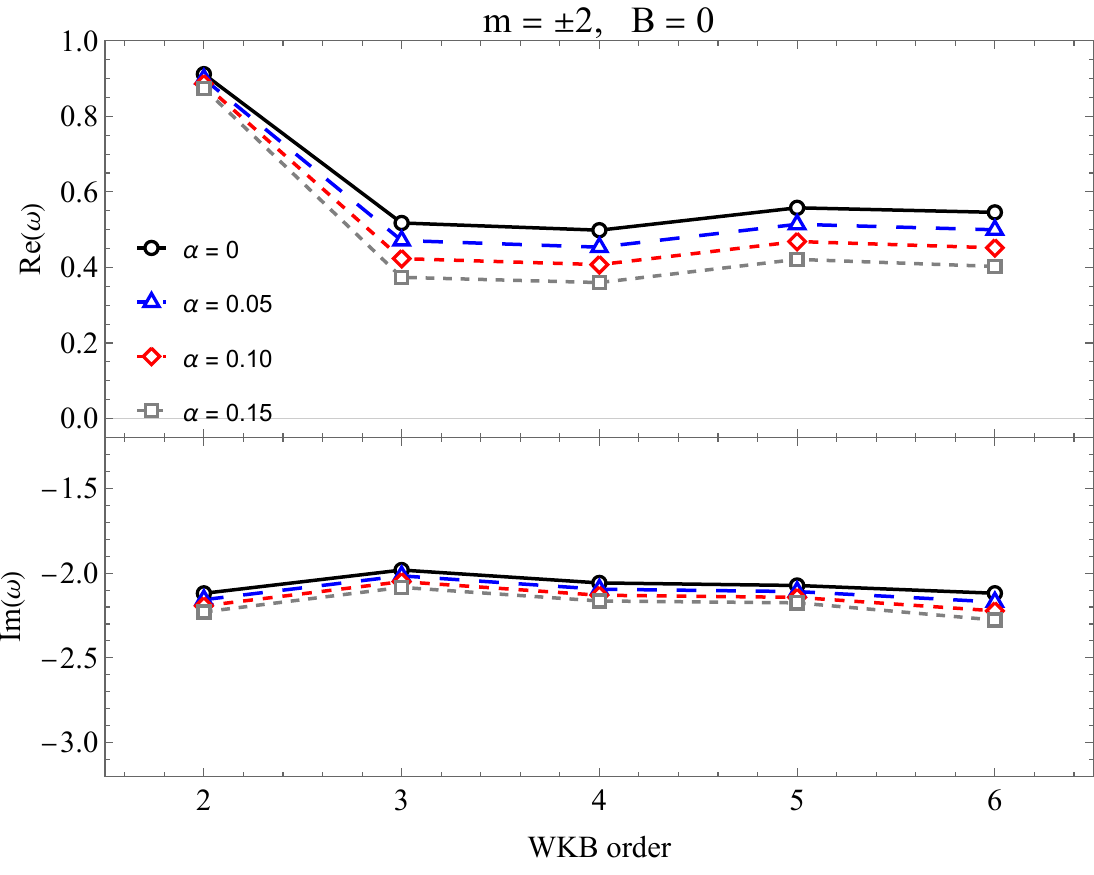}\label{WKBorder_m2B0n2}}
 \caption{\footnotesize{Real and imaginary parts of the quasinormal frequency as a function of the WKB order for $m = \pm1,\pm2$ and overtone number $n=2$.}}
  \label{WKBorder}
 \end{figure}

\begin{figure}[htbp]
 \centering
\subfigure[]{\includegraphics[width=.48\textwidth]{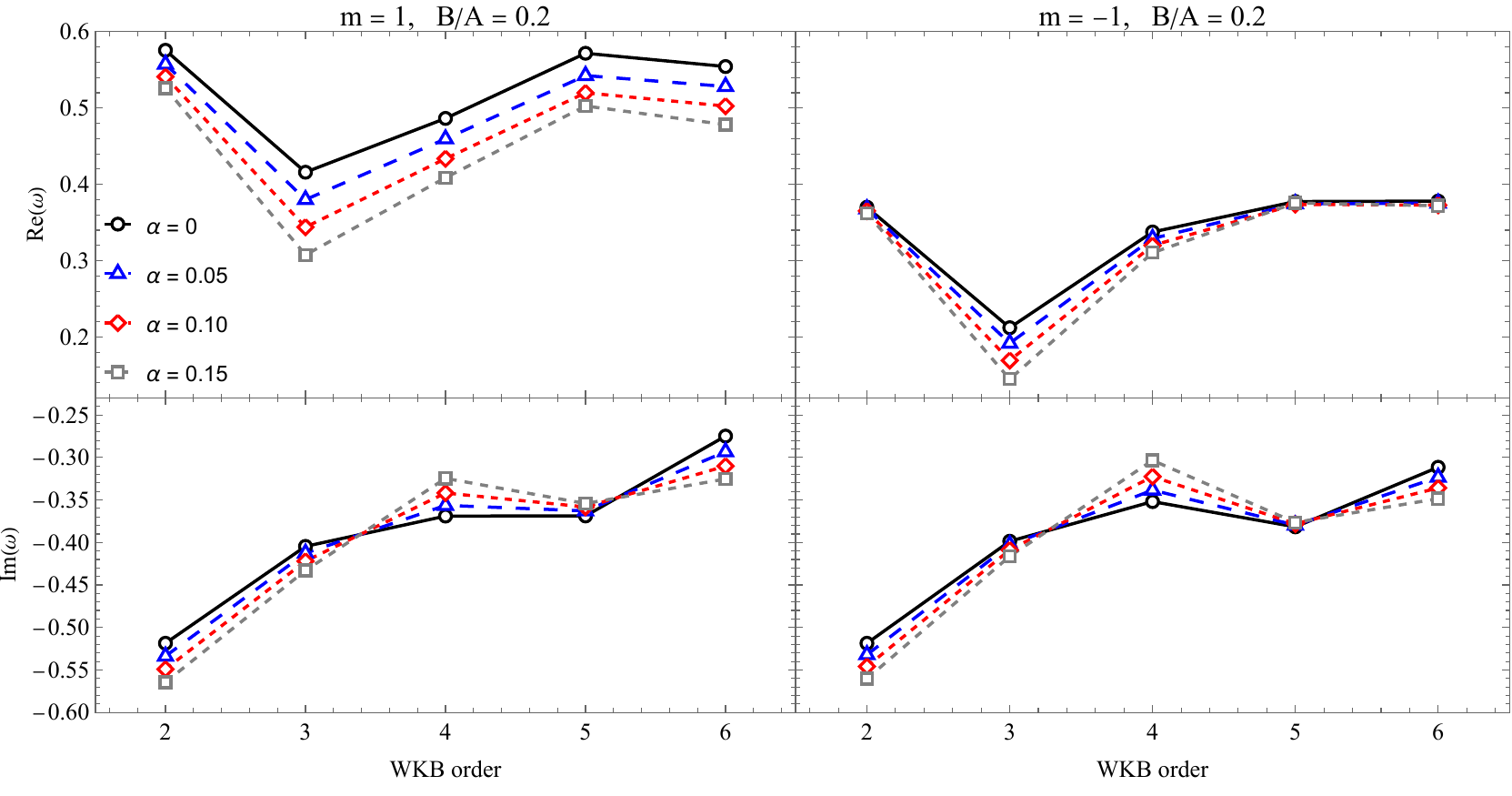}\label{WKBorder_m1B02}}
\quad
\subfigure[]{\includegraphics[width=.48\textwidth]{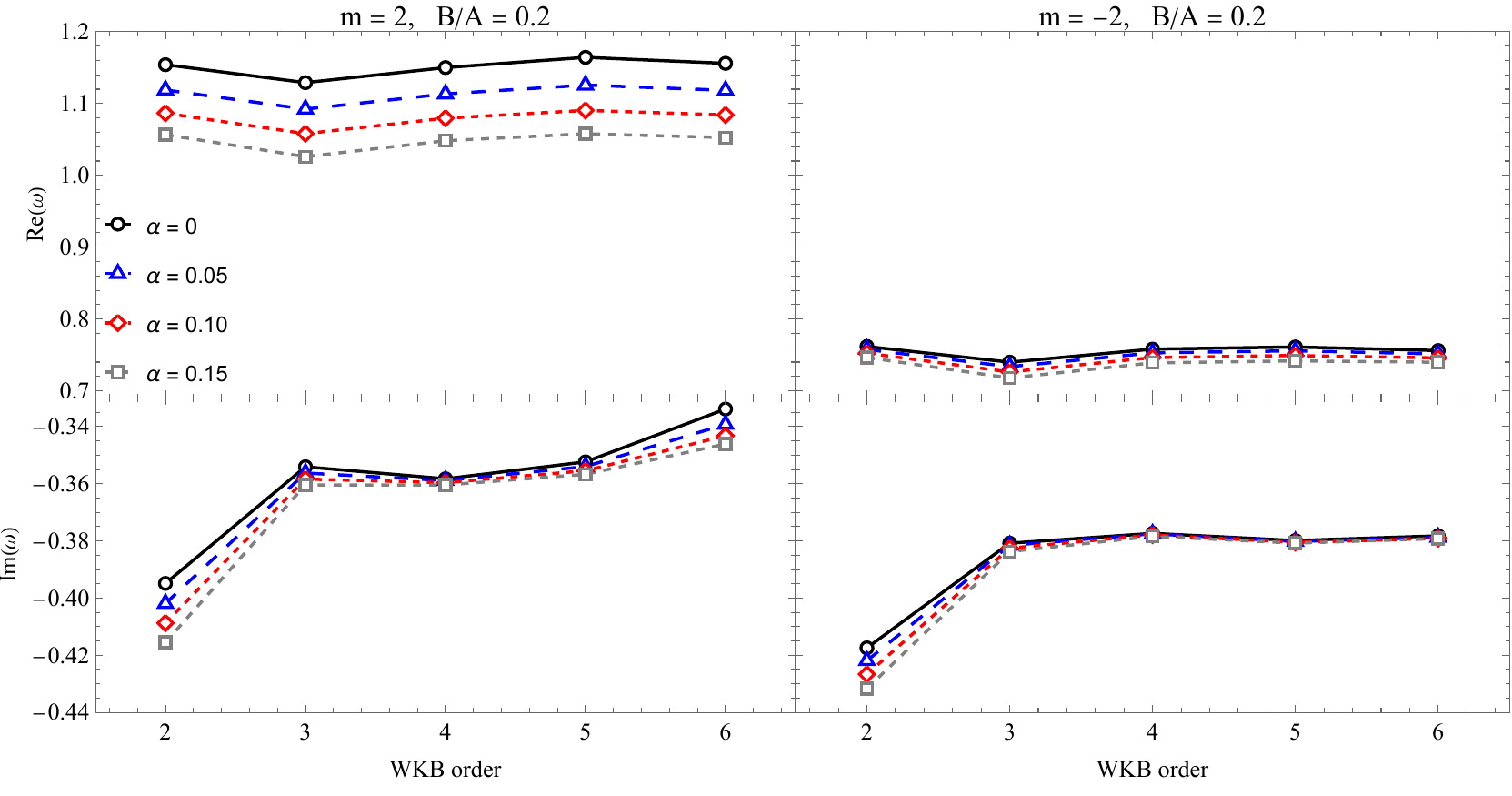}\label{WKBorder_m2B02}}
  \caption{\footnotesize{Real and imaginary parts of the quasinormal frequency as a function of the WKB order for $m = \pm1,\pm2$ and overtone number $n=0$ and $B/A=0.2$.}}
  \label{WKBorderB02}
 \end{figure}

Figures \ref{WKBorder} and \ref{WKBorderB02} show the results of the quasinormal frequency, real and imaginary parts, for orders $2$ to $6$ of the WKB approximation for $m = \pm1, \pm2$. Figure \ref{WKBorder} presents the results for the non-rotating case $B=0$ with overtone number $n=2$, while Fig. \ref{WKBorderB02} presents the results for the rotating case $B/A=0.2$ with $n=0$. For $m=\pm2$, the results remain stable with an increasing order of approximation in both cases. For $m=\pm1$ with rotation (Fig. \ref{WKBorder_m1B02}), the results also exhibit good stability. However, for $m=\pm 1$ without rotation (Fig. \ref{WKBorder_m1B0n2}) there is instability for lower orders of the WKB approximation.
The results obtained for the quasinormal frequencies ($A\omega_{n}$) are presented in Tables \ref{tab1} and \ref{tab2}. In Table \ref{tab1}, we display the spectrum for the case without rotation ($B=0$) as a function of the Lorentz breaking parameter $\alpha$, for different azimuthal modes $m$ and overtone numbers $n$.
It can be observed that for most modes increasing $\alpha$ systematically causes a reduction in the real part of the frequency and an increase in the magnitude of its imaginary part, indicating that Lorentz breaking makes the oscillations more damped.  
For $\alpha=0$, the results reproduce those found in the literature for the conventional acoustic black hole~\cite{cardoso2004quasinormal,berti2004quasinormal}. 
As a benchmark, we compare our WKB results for the fundamental mode ($n=0$) with those reported in Table I of~\cite{berti2004quasinormal}. For $m=1,2$ and $3$, our results $\omega_{0} = 0.42722 - 0.33011i, 0.95143 - 0.35304i$ and $1.46852 - 0.35248i$, agree with the values $0.427 - 0.330i, 0.951 - 0.353i$ and $1.468-0.353i$ from Ref.~\cite{berti2004quasinormal}, respectively.

Table \ref{tab2} shows the combined effect of rotation ($B$) and Lorentz breaking ($\alpha$) for the fundamental mode ($n=0$). The asymmetry between co-rotating ($m>0$) and counter-rotating ($m<0$) modes is evident, reflecting the influence of rotation on the spectrum. The effect of Lorentz symmetry breaking continues to reduce the real frequency part and increase the imaginary part. However, as the rotation increases, the influence of the parameter $\alpha$ on the imaginary part is attenuated for counter-rotating modes, as can be observed for the case $m=-2$ with $B/A=0.2$. This indicates that, for higher values of rotation, the effects of Lorentz symmetry breaking do not significantly affect the damping rate in counter-rotating scenarios.
\begin{table}[h!]
	\begin{footnotesize}
	\begin{center}
	\caption{\footnotesize{Quasinormal frequencies $\omega A$ ($B = 0$).}} 
	\label{tab1}	
	\begin{tabular}{|c||c||c|c|c|c|c}
	\hline
 $ \alpha $ & $n$   &  $m = 1$ &   $m = 2$ &  $m = 3$ &  $m = 4$  \\
	\hline
	\multirow{3}{*}{$ 0 $}
   & 0  & 0.42722 - 0.33011i & 0.95143 - 0.35304i & 1.46852 - 0.35248i & 1.97645 - 0.35296i   \\
   & 1  & 0.18880 - 1.18195i & 0.78348 - 1.13492i & 1.34827 - 1.08979i & 1.88455 - 1.07683i   \\
   & 2  & 0.56936 - 2.09125i & 0.54575 - 2.11877i & 1.13693 - 1.92391i & 1.71231 - 1.85554i   \\
	\hline
	\hline
	\multirow{3}{*}{$ 0.05 $}
   & 0  & 0.41615 - 0.33501i & 0.93757 - 0.35647i & 1.45072 - 0.35622i & 1.95402 - 0.35686i   \\
   & 1  & 0.16040 - 1.19454i & 0.75794 - 1.15139i & 1.32243 - 1.10278i & 1.85610 - 1.08948i   \\
   & 2  & 0.57099 - 2.17802i & 0.49965 - 2.17016i & 1.09499 - 1.95353i & 1.67163 - 1.88051i   \\
	\hline
	\hline
	\multirow{3}{*}{$ 0.10 $}
   & 0  & 0.40534 - 0.34065i & 0.92382 - 0.35944i & 1.43294 - 0.35966i & 1.93161 - 0.36052i   \\
   & 1  & 0.13045 - 1.20389i & 0.73203 - 1.16699i & 1.29637 - 1.11482i & 1.82751 - 1.10134i   \\
   & 2  & 0.56964 - 2.28287i & 0.45186 - 2.22257i & 1.05192 - 1.98181i & 1.63029 - 1.90419i   \\
	\hline
	\hline
	\multirow{3}{*}{$ 0.15 $}
   & 0  & 0.395353 - 0.34677i & 0.910334 - 0.36194i & 1.41530 - 0.36286i & 1.90940 - 0.36396i   \\
   & 1  & 0.098201 - 1.20806i & 0.705857 - 1.18187i & 1.27021 - 1.12597i & 1.79893 - 1.11253i   \\
   & 2  & 0.565402 - 2.40860i & 0.402611 - 2.27665i & 1.00777 - 2.00884i & 1.58840 - 1.92675i   \\
	\hline
 	\end{tabular}
	\end{center}
	\end{footnotesize}
\end{table}

\begin{figure}[!htb]
 \centering
 \subfigure[]{\includegraphics[scale=0.35]{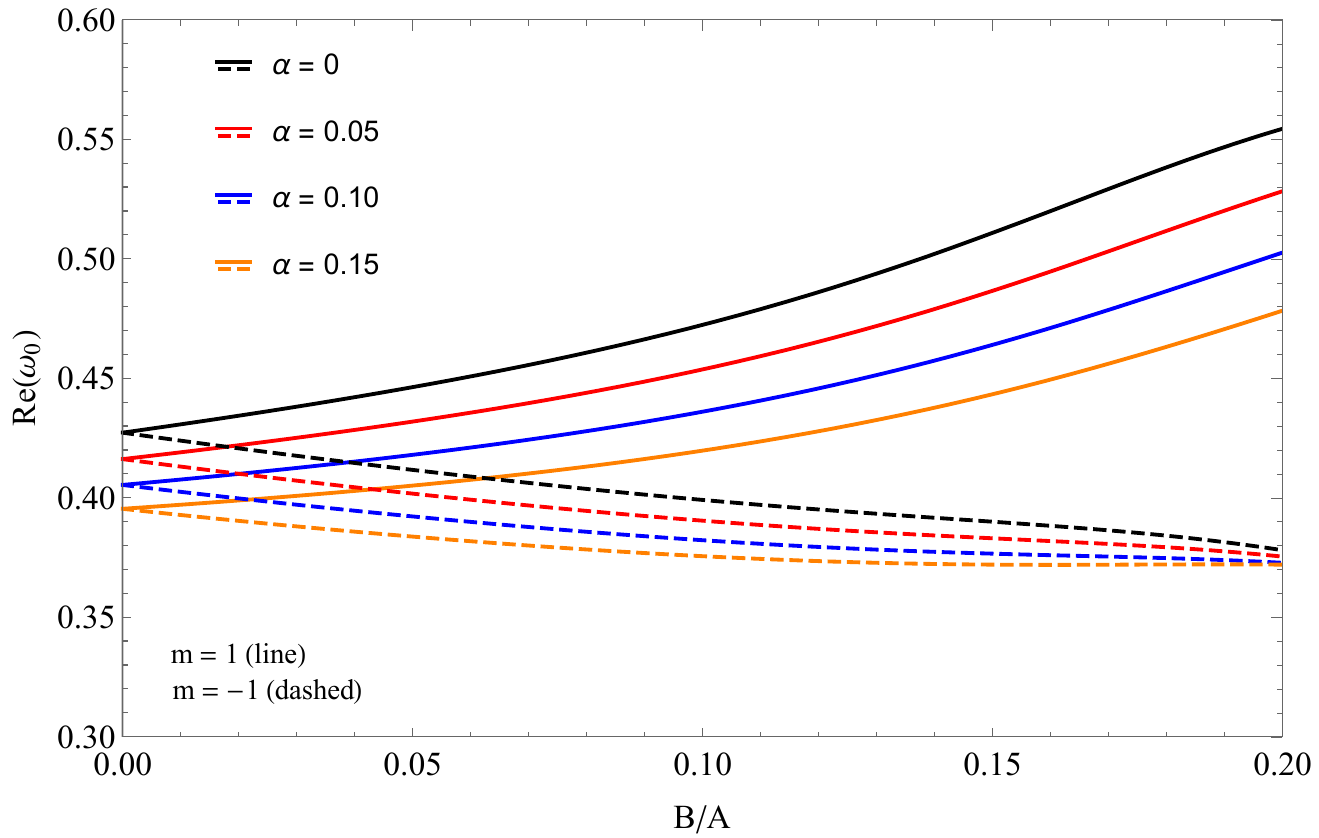}\label{Re_w0_m1}}
 \quad
 \subfigure[]{\includegraphics[scale=0.35]{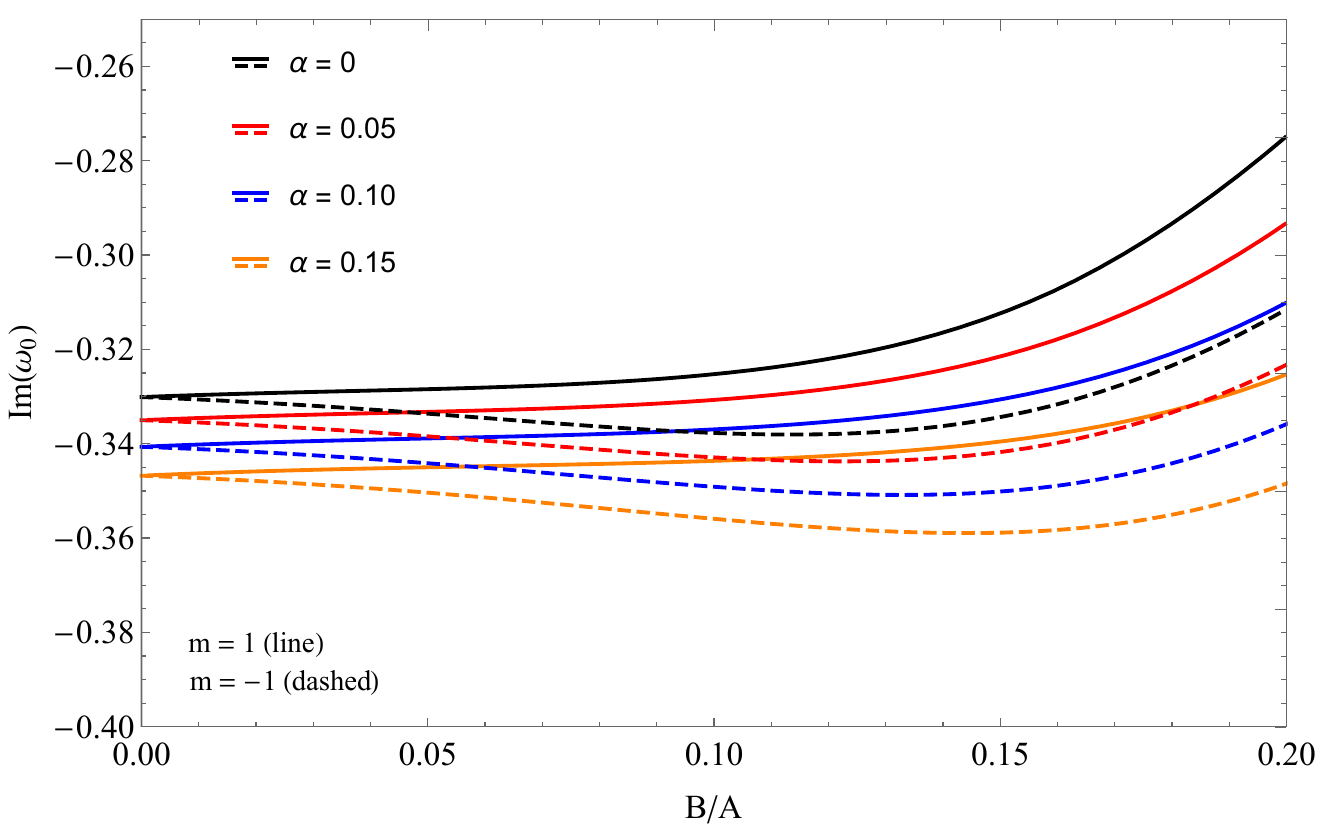}\label{Im_w0_m1}}
 \quad
 \subfigure[]{\includegraphics[scale=0.35]{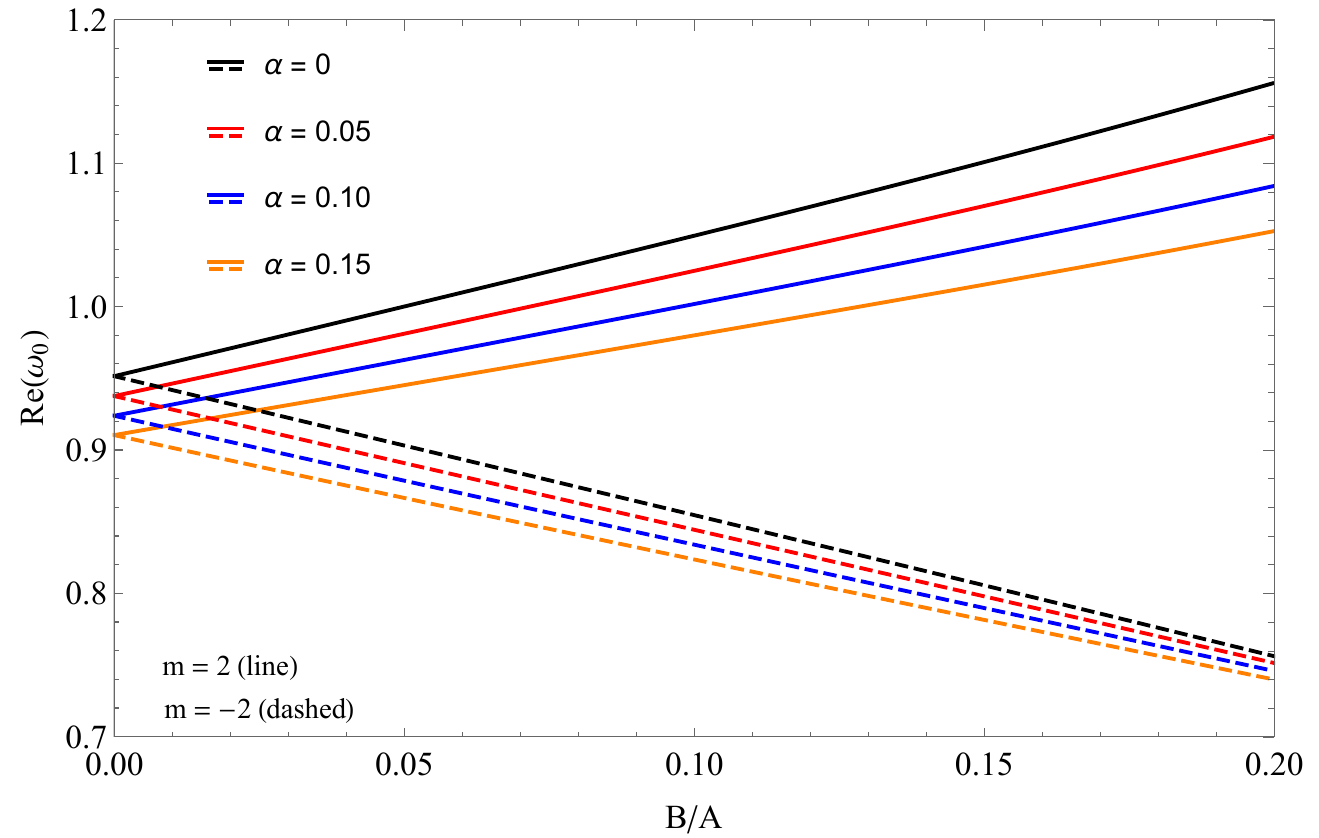}\label{Re_w0_m2}}
 \quad
 \subfigure[]{\includegraphics[scale=0.35]{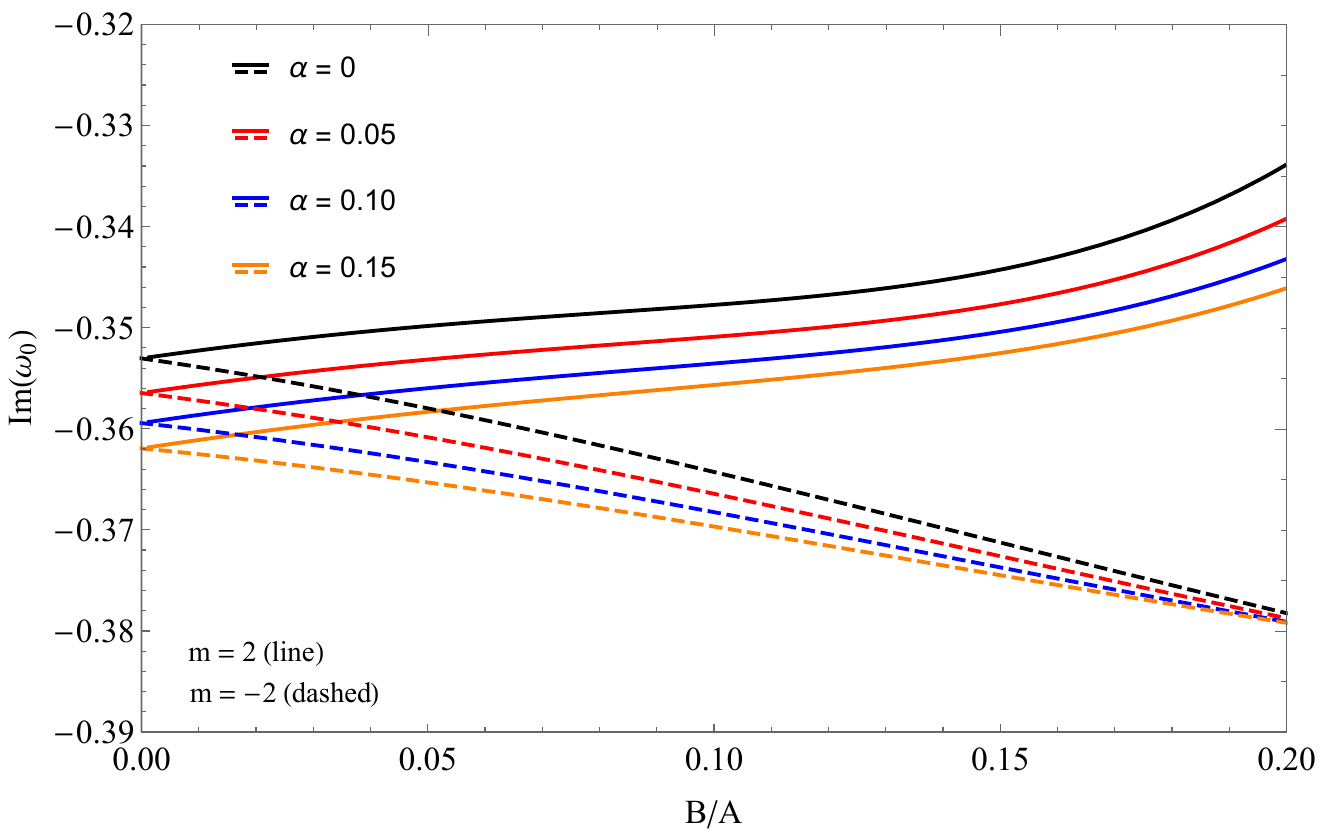}\label{Im_w0_m2}}
  \caption{\footnotesize{Real (left) and imaginary (right) parts of the fundamental quasinormal frequency ($n=0$) as a function of the rotation parameter $B$, for $m=1$ (upper panels) and $m=2$ (lower panels). Solid (dashed) lines represent modes with $m>0$ ($m<0$). Different colors indicate different values of the Lorentz breaking parameter $\alpha$.}}
 \label{Re_Im_w0_m1_m2}
\end{figure}

\begin{table}[h!]
	\begin{footnotesize}
	\begin{center}
	\caption{\footnotesize{Quasinormal frequencies $\omega_{n} A$ ($n = 0$).}} 
	\label{tab2}	
	\begin{tabular}{|c||c||c|c|c|c}
	\hline
 $ m $ & $\alpha$   &  $B/A = 0.05$ &   $B/A = 0.10$ &  $B/A = 0.20$  \\
	\hline
	\multirow{4}{*}{$ 1 $}
   & 0    & 0.446265 - 0.328437i & 0.472321 - 0.325254i & 0.554392 - 0.274826i    \\
   & 0.05 & 0.431847 - 0.333247i & 0.453676 - 0.330713i & 0.528251 - 0.293227i    \\
   & 0.10 & 0.417936 - 0.338875i & 0.435985 - 0.336961i & 0.502581 - 0.310071i    \\
   & 0.15 & 0.405063 - 0.344994i & 0.419692 - 0.343629i & 0.478267 - 0.325298i    \\
	\hline
	\hline
	\multirow{4}{*}{$ -1 $}
   & 0    & 0.411644 - 0.333620i & 0.399071 - 0.337711i & 0.378031 - 0.311489i    \\
   & 0.05 & 0.401713 - 0.338424i & 0.390387 - 0.342893i & 0.375345 - 0.323245i    \\
   & 0.10 & 0.392163 - 0.344128i & 0.382212 - 0.349121i & 0.372764 - 0.335846i    \\
   & 0.15 & 0.383730 - 0.350367i & 0.375523 - 0.355913i & 0.371983 - 0.348446i    \\
	\hline
	\hline
	\multirow{4}{*}{$ 2 $}
   & 0     & 1.000000 - 0.349833i & 1.049380 - 0.347749i & 1.15598 - 0.333865i    \\
   & 0.05  & 0.981034 - 0.353151i & 1.024890 - 0.350905i & 1.11844 - 0.339221i    \\
   & 0.10  & 0.962740 - 0.355984i & 1.001740 - 0.353547i & 1.08412 - 0.343202i    \\
   & 0.15  & 0.945187 - 0.358324i & 0.979905 - 0.355668i & 1.05263 - 0.346095i    \\
	\hline
	\hline
	\multirow{4}{*}{$ -2 $}
   & 0     & 0.903072 - 0.357975i & 0.854529 - 0.364275i & 0.756114 - 0.378237i    \\
   & 0.05  & 0.890804 - 0.360840i & 0.844317 - 0.366431i & 0.751371 - 0.378733i    \\
   & 0.10  & 0.878555 - 0.363294i & 0.833936 - 0.368248i & 0.745808 - 0.379099i    \\
   & 0.15  & 0.866495 - 0.365307i & 0.823609 - 0.369669i & 0.739838 - 0.379193i    \\
	\hline
 	\end{tabular}
	\end{center}
	\end{footnotesize}
\end{table}
Figure \ref{Re_Im_w0_m1_m2} illustrates the behavior of the real and imaginary parts of the frequencies for $m = \pm 1$ and $m = \pm 2$ as a function of the rotation parameter $B$, for different values of $\alpha$. It can be observed that, for co-rotating modes ($m>0$, solid lines), increasing the rotation intensifies the effect of the parameter $\alpha$, especially in the imaginary part of $m=1$. On the other hand, for counter-rotating modes ($m<0$, dashed lines), with the exception of the imaginary part of $m=-1$, the curves for different values of $\alpha$ converge when $B/A=0.2$, indicating that, in this regime, rotation dominates the dynamics and the effect of Lorentz breaking is suppressed.

\section{Conclusions}
\label{S5}
In this work, we investigated the effects of Lorentz symmetry violation on the absorption cross section and quasinormal modes of a rotating acoustic black hole in (2+1) dimensions within the regime of slow rotation. The metric is derived from the Abelian Higgs model through the inclusion of a Lorentz-violating parameter $\alpha$, and describes a sonic analogue of a rotating black hole.
The results demonstrate, under these approximations that Lorentz violation significantly modifies the absorption and emission dynamics of scalar waves in this system. In the low-frequency regime, an analytical analysis revealed that the absorption cross section receives an explicit contribution from the rotation parameter $B$, which is dependent on the presence of $\alpha$. In the high-energy limit, a geodesic study showed that the critical impact parameter, and consequently the classical absorption cross section, are also increased by the Lorentz violation term. The numerical solution of the radial equation confirmed these asymptotic behaviors and provided the absorption cross section for the entire frequency spectrum, validating the consistency of our analyses within the restricted parameter space. We analyzed quasinormal modes using a sixth-order WKB approximation, and verified that the presence of the parameter $\alpha$ causes a systematic reduction in the real part of the frequencies and an increase in the magnitude of their imaginary part. This indicates that Lorentz breaking makes the field oscillations more damped. We also observed that, as the rotation $B$ increases, the contribution of $\alpha$ increases for the co-rotating modes ($m>0$), evidencing a coupling between the rotation of the acoustic black hole and the symmetry violation term.

Finally, our results indicate that the Lorentz symmetry violation parameter increase the effective radius of the capture region (analogous to the shadow of the black hole), which is directly related to the observed increase in the absorption cross section across the entire frequency spectrum is important to emphasize that all conclusions are valid strictly within the slow rotation and small $\alpha$. This work, therefore, contributes to the understanding of how fundamental symmetry violations can affect analog gravity models.

{\acknowledgments We thank CNPq and CAPES for partial financial support. M.A.A, F.A.B, E.P and A.R.Q acknowledges support from CNPq (Grant nos. $301683/2025-5$, $309092/2022-1$, $304290/2020-3$ and $ 310533/2022-8$). JAVC thanks the Paraíba State Research Support Foundation (FAPESQ) (Grant No. $22/2025$) for financial support. }

\bibliographystyle{ieeetr}

\end{document}